\documentclass[conference]{IEEEtran}
\IEEEoverridecommandlockouts
\usepackage{cite}
\usepackage{amsmath,amssymb,amsfonts}
\usepackage{amsthm}
\newtheorem{definition}{Definition}

\usepackage[ruled,vlined,linesnumbered]{algorithm2e}

\makeatletter
\newcounter{functioncf}

\newenvironment{FunctionInline}{
  \refstepcounter{functioncf}%
  \@twocolumnfalse
  \begin{algorithm}[H]
}{
  \end{algorithm}
}
\makeatother
\usepackage{url}
\usepackage{algpseudocode}
\usepackage{graphicx}
\usepackage{textcomp}
\usepackage[svgnames]{xcolor}
\usepackage{enumitem}
\usepackage{multirow}
\usepackage{booktabs}
\newtheorem{theorem}{Theorem}
\usepackage{svg}
\usepackage{arydshln}
\usepackage{subcaption}

\def\BibTeX{{\rm B\kern-.05em{\sc i\kern-.025em b}\kern-.08em
    T\kern-.1667em\lower.7ex\hbox{E}\kern-.125emX}}
\begin{document}

\title{CARROT: A Learned Cost-Constrained Retrieval Optimization System for RAG\\

}

\author{
\IEEEauthorblockN{
Ziting Wang\IEEEauthorrefmark{2},
Haitao Yuan\IEEEauthorrefmark{2}\thanks{Corresponding author: Haitao Yuan (haitao.yuan@ntu.edu.sg)},
Wei Dong\IEEEauthorrefmark{2},
Gao Cong\IEEEauthorrefmark{2},
Feifei Li\IEEEauthorrefmark{3}
}
\IEEEauthorblockA{
\IEEEauthorrefmark{2}\textit{Nanyang Technological University, Singapore}
}
\IEEEauthorblockA{
\IEEEauthorrefmark{3}\textit{Alibaba Group, China}
}
\IEEEauthorblockA{
\IEEEauthorrefmark{2}ziting001@e.ntu.edu.sg, \IEEEauthorrefmark{2}\{haitao.yuan, wei\_dong, gaocong\}@ntu.edu.sg, \IEEEauthorrefmark{3}lifeifei@alibaba-inc.com
}
}

\maketitle

\begin{abstract}
Large Language Models (LLMs) have demonstrated impressive ability in generation and reasoning tasks but struggle with handling up-to-date knowledge, leading to inaccuracies or hallucinations. Retrieval-Augmented Generation (RAG) mitigates this by retrieving and incorporating external knowledge into input prompts. In particular, due to LLMs' context window limitations and long-context hallucinations, only the most relevant ``chunks'' are retrieved. However, current RAG systems face three key challenges: (1) chunks are often retrieved independently without considering their relationships, such as redundancy and ordering; (2) the utility of chunks is non-monotonic, as adding more chunks can degrade quality; and (3) retrieval strategies fail to adapt to the unique characteristics of different queries.

To overcome these challenges, we design a cost-constrained retrieval optimization framework for RAG. We adopt a Monte Carlo Tree Search (MCTS) based strategy to find the optimal chunk combination order, which considers the chunks' correlations. In addition, to address the non-monotonicity of chunk utility, instead of treating budget exhaustion as the termination condition, we design a utility computation strategy to identify the optimal chunk combination without necessarily exhausting the budget. Furthermore, we propose a configuration agent that predicts optimal configurations for each query domain, improving our framework's adaptability and efficiency. Experimental results demonstrate up to a 30\% improvement over baseline models, highlighting the framework's effectiveness, scalability, and suitability. Our source code has been released at \url{https://github.com/wang0702/CARROT}.

\end{abstract}
    \section{Introduction\label{sec:intro}}
Although LLMs have demonstrated exceptional capabilities in generative tasks, they still struggle with accessing up-to-date information~\cite{chat2data,DBLP:journals/pvldb/AutoTQA,DBLP:journals/pacmmod/tablegpt,DBLP:journals/pvldb/LLMdatalake}, leading to hallucinations~\cite{hall}. RAG has emerged as a crucial solution to address this issue, especially for tasks requiring updated external knowledge such as question answering tasks~\cite{hall,raptor,NL2SQL,DBLP:journals/pvldb/AutoTQA}, personalized agents~\cite{crafting}, and time series analysis~\cite{DBLP:conf/emnlp/FonsKPZBVV24,DBLP:journals/pvldb/tsg}. External data is often too large to input directly into LLMs, so it is split into chunks stored in a vector database~\cite{DBLP:journals/pacmmod/debugrag,richrag,GraphRAG,DBLP:conf/nips/rankrag,DBLP:conf/acl/Mrag}. Users query these chunks to create prompts, making efficient and accurate search algorithms a key research focus in the information retrieval domain~\cite{hall,DBLP:journals/pvldb/text2sqlnew,DBLP:journals/pacmmod/tablegpt,DBLP:journals/pvldb/AutoTQA,DBLP:journals/pvldb/LLMdatalake,chat2data}.
Note that although Agentic RAG~\cite{DBLP:conf/emnlp/FLARE,DBLP:journals/corr/abs-2501-09136} establishes a new paradigm for complex, multi-step tasks, it does not render standard RAG obsolete and may rely on standard RAG for subtasks. Moreover, standard RAG remains the default choice for efficient, single-turn information retrieval in enterprise Q\&A, as evidenced by results on knowledge-intensive tasks and broad industry adoption~\cite{DBLP:conf/nips/CRAG, DBLP:journals/pvldb/vectordb}.

However,
many existing RAG frameworks incur expensive computational costs. First, tuning-based RAG approaches~\cite{DBLP:conf/nips/rankrag, DBLP:conf/iclr/selfrag, DBLP:conf/naacl/tuning, DBLP:conf/sigir/colbert, DBLP:conf/naacl/colbert2} require extensive instruction tuning to enhance LLMs' abilities in context evaluation and utilization. These methods fine-tune models on specialized tasks like context ranking~\cite{DBLP:conf/nips/rankrag} or self-reflection~\cite{DBLP:conf/iclr/selfrag} to improve retrieval quality and generation accuracy. While effective, they demand significant computational cost for model training and specialized datasets. Similarly, graph-based RAG methods~\cite{GraphRAG, DBLP:conf/acl/JinXZRZL0TWM024, gretriever, raptor, DBLP:conf/nips/hipporag1} rely on LLM inference to construct knowledge graphs or guide actions. While these graph-based approaches avoid tuning models, they impose significant computational overhead by requiring multiple LLM inferences for graph construction and chunk summarization, making them prohibitively expensive when scaling to large knowledge bases.

\begin{figure}[t]
  \centering
  \includegraphics[width=\columnwidth]{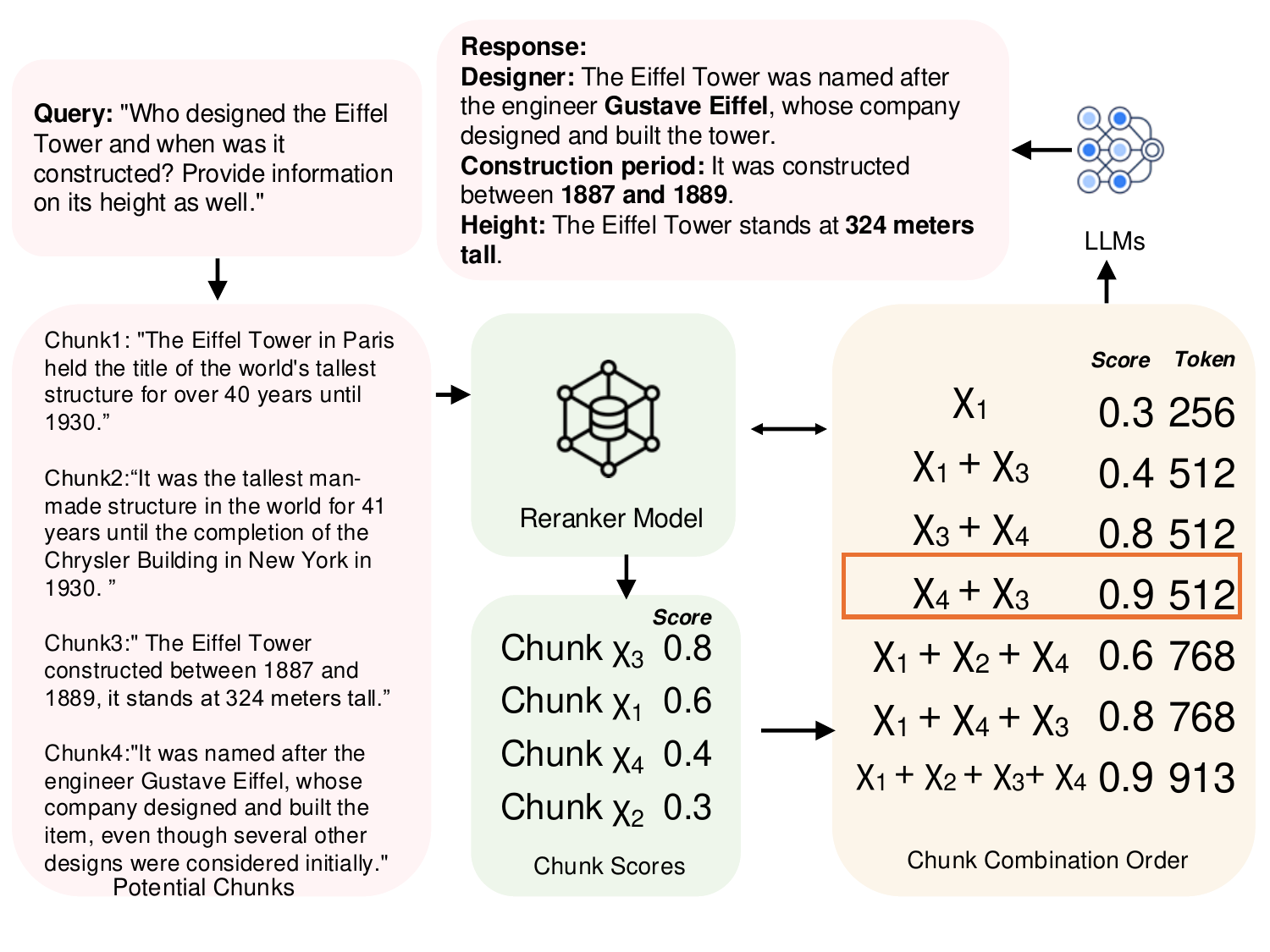}
  \vspace{-20pt}
  \caption{Chunk combination order example.}
  \label{fig:example}
  \vspace{-10pt}
\end{figure}

\begin{figure*}[!t]
  \centering
  \includegraphics[width=0.92\textwidth]{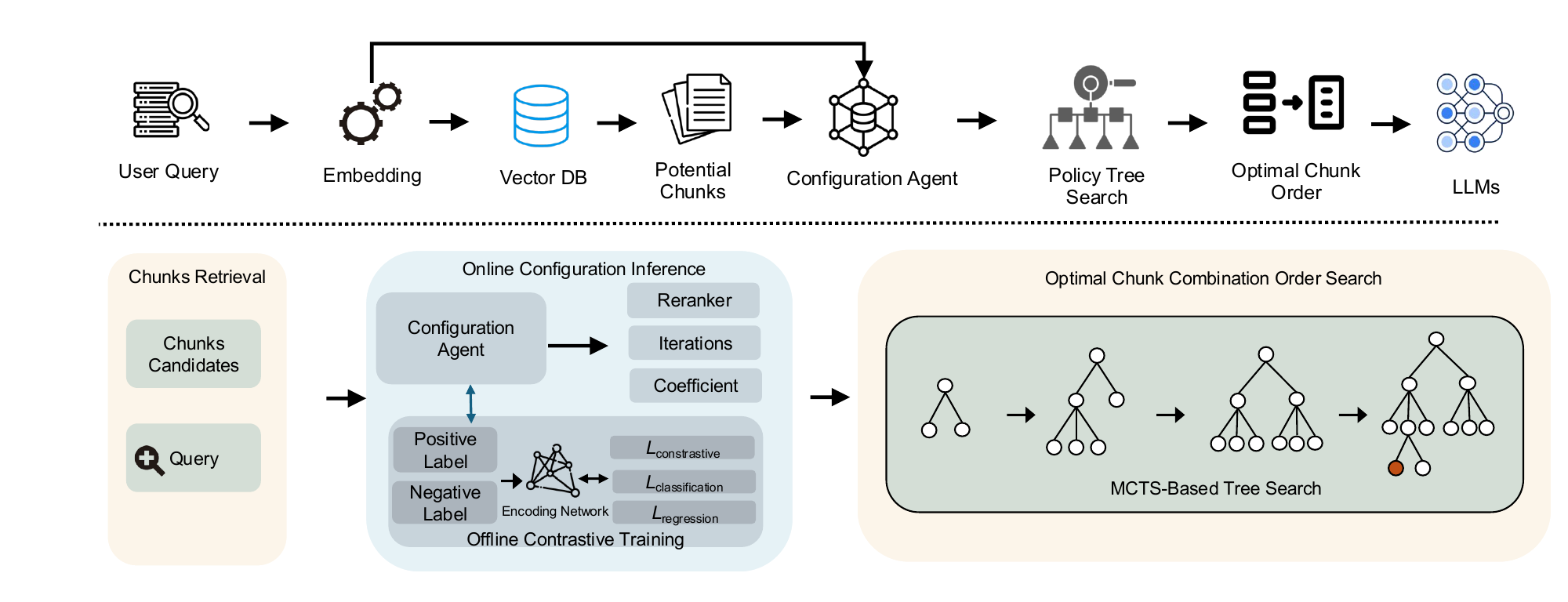}
  \vspace{-10pt}
  \caption{The architecture overview of CARROT framework.}
  \label{fig:framework}
  \vspace{-10pt}
\end{figure*}

Besides tuning-based and graph-based methods, rank-based RAG represents another prevalent approach~\cite{naiverag, DBLP:conf/acl/Mrag,DBLP:conf/acl/hyde, richrag}. These methods assign a utility score to each chunk, indicating its relevance for answering the user query. These scores typically derive from similarity metrics between queries and chunks or from specialized reranker models~\cite{RL_reranker, naiverag, DBLP:conf/aaai/Retriever-Reranker}, whose scores have demonstrated strong correlation with final response quality in production deployments~\cite{llamaindex, DBLP:journals/corr/jinarerankerv3}. Unlike tuning-based and graph-based methods that require substantial LLM modifications and interactions, rank-based methods usually employ reranker models, offering reasonable computational efficiency. However, our empirical study in Table~\ref{tab:rouge_comparison} reveals that existing rank-based methods underperform in
difficult query workloads and long-context scenarios, presenting three key challenges.

\textit{Challenge 1: Relationships between chunks.}
There are two main methods to identify relevant chunks. The first uses approximate k-nearest neighbor (AKNN) to rank and select top-$k$ chunks~\cite{DBLP:journals/pvldb/AKNN}. The second clusters chunks and returns all chunks in the most relevant clusters~\cite{raptor,GraphRAG}. However, AKNN disregards correlations entirely, and clustering treats all chunks in a cluster as equally relevant, leading to redundancy when chunks have overlapping information. For example, in Figure~\ref{fig:example}, querying the height and history of the Eiffel Tower using the AKNN method would select \(\chi_3\) and \(\chi_1\), which only provide historical details, failing to address the query fully. However, the clustering approach returns all chunks (\(\chi_1, \chi_2, \chi_3, \chi_4\)), causing redundancy. The optimal solution is to select \(\chi_3\) and \(\chi_4\), avoiding redundancy while meeting the query's intent.  In addition, previous work~\cite{lostmiddle, DBLP:conf/acl/order} demonstrates that information position significantly affects RAG performance, a phenomenon known as ``lost in the middle''~\cite{lostmiddle}. For instance, placing \(\chi_4\) before \(\chi_3\) yields better results. However, finding the best chunk combination and order is challenging due to the exponentially growing search space, and we prove this problem is NP-hard in Section~\ref{sec:prelimi}.

\textit{Challenge 2: Non-monotonicity of utility.\label{mono}}
Current solutions
are typically  based on the assumption that including more chunks will
yield better final results. Specifically, in the AKNN-based approach, exactly $k$ chunks are selected
each time. In the clustering approach, a distance threshold between clusters and the query is fixed, and all clusters within this threshold are returned.
Both of them return as many chunks as possible. However, in practice, the utility of chunks is not monotonic.
More specifically, excessive chunks can dilute key information by adding marginally relevant content, creating noise that reduces clarity. Additionally, conflicting or nuanced differences across chunks may confuse the model, leading to hallucinations.
For example, as illustrated in Figure \ref{fig:example}, when $\chi_3$ and $\chi_4$ are selected, adding chunk $\chi_1$ decreases utility, highlighting that utility scores are often non-monotonic.

\textit{Challenge 3: Diversity of queries.}
Users' queries have different types, requiring different ranking strategies due to their unique characteristics~\cite{DBLP:journals/pvldb/diver, DBLP:conf/acl/Mrag}.
Current RAG systems predominantly rely on a single reranker model~\cite{RL_reranker,DBLP:conf/acl/lightrerank} to determine chunks' utility scores.
Although various reranker models have been developed, their performance varies significantly by query types, with no single model consistently outperforming others across all variations
(as shown in our preliminary experiments in Fig.~\ref{fig:Distribution}).
Current RAG systems~\cite{richrag, naiverag}
lack the flexibility to adapt to varying query intentions.

\noindent \textbf{Problem Statement.~}
\textit{Is there a RAG framework that achieves both high efficacy and efficiency while maintaining acceptable computational costs?}

\noindent \textbf{Our contributions.}
We answer this question by proposing a novel rank-based RAG framework named \textbf{\underline{c}}ost-constr\textbf{{\underline{a}}}ined \textbf{\underline{r}}et\textbf{\underline{r}}ieval \textbf{\underline{o}}p\textbf{\underline{t}}imization (\textit{CARROT}).
On one hand, \textit{CARROT} is a rank-based framework, thus having much higher efficiency than graph-based and tuning-based methods. On the other hand, it considers chunks correlation, non-monotonicity, and query intent diversity in the search process to enhance response quality. Additionally, to further reduce computational costs and make the framework practical, we minimize token cost by incorporating cost constraints in the
search process.

\begin{itemize}[leftmargin=10.2pt,noitemsep,topsep=0pt]
\setlength{\itemsep}{0pt}
\setlength{\parsep}{0pt}
\setlength{\parskip}{0pt}
\item We propose the first RAG framework that explicitly considers chunk combination order for the RAG task. Rather than treating each chunk independently or at the cluster level, we design an MCTS-based policy tree search to identify the optimal chunk combination order. Unlike greedy methods that
select the highest-scored chunk at each step, our approach iteratively samples and evaluates different paths, leveraging a UCB-based utility function to balance exploring new combinations with exploiting promising ones. This design addresses the exponential search space by focusing computational resources on high-potential paths, while our parallel evaluation strategy scores multiple candidates simultaneously for efficiency. Through these designs, we address \textit{Challenge 1} while maintaining computational efficiency.

\item In contrast to existing RAG frameworks that treat budget exhaustion merely as a termination criterion, we design a novel formulation that directly incorporates budget constraints into the optimization of chunk combinations. Our formulation explicitly captures the non-monotonic utility of chunks, effectively addressing \textit{Challenge 2}.

\item  We propose a contrastive learning-based configuration agent that dynamically predicts MCTS configurations per query, adapting reranker models and configurations to the specific query domain. This approach tailors retrieval for dynamic, domain-specific queries with flexibility and robustness, addressing \textit{Challenge 3}.

\item We conducted comprehensive experiments comparing our framework against both graph-based and rank-based methods. The results validate the effectiveness, efficiency, and scalability of our approach, demonstrating a 30\% performance improvement over baseline methods.

\end{itemize}

    \section{Preliminaries\label{sec:prelimi}}

We first introduce the definitions of chunks and chunk combination orders.
We then formalize the chunk order optimization problem and give the NP-hard proof of the problem.

\noindent \textbf{RAG \& Chunks.} RAG improves generation models by retrieving relevant context from external corpora, which is first divided into smaller, manageable units called \textit{chunks}, typically stored in a vector database.
The definition of chunk is given as follows:
\vspace{-0.05in}
\begin{definition}[Chunk]
Let \( C \) represent a corpus of documents, and a chunk \( \chi \) is defined as a contiguous block of text extracted from \( C \). Formally, a chunk \( \chi \) consists of a sequence of tokens \( (t_1, t_2, \ldots, t_n) \), where each \( t_i \) is a token from \( C \) and the size \( n \) is set by users.
\end{definition}
\vspace{-0.05in}

\noindent \textbf{Chunk Cost \& Benefit.} Similar to prior work~\cite{doctopus}, the cost of a chunk is defined as \(\text{cost}(\chi) = |\chi|\), representing the number of tokens it contains; thus, longer chunks incur higher costs. The benefit of a chunk reflects its estimated contribution to the final response quality. In RAG systems, each chunk is embedded into a vector representation using an embedding model to capture its semantic information. Given a query \(Q\), the vector database performs a similarity search to identify the most relevant chunks, which are then passed to the generator (e.g., a large language model) to produce the final response. In this phase, we measure the benefit of a chunk by its estimated contribution to the final response quality. We measure this using a reranker model: \(W(Q,\chi) = \text{Reranker}(Q, \chi)\), where the reranker is a neural model trained to predict the relevance between query \(Q\) and chunk \(\chi\). For simplicity, we use \(W(Q, \chi)\) and \(W(\chi)\) interchangeably when no ambiguity arises.

\noindent \textbf{Chunk Combination Order.}
In an RAG system, retrieval from the vector database often yields multiple chunks. However, due to input constraints and hallucination of the generation model, incorporating all of them is infeasible. Thus, it is necessary to select an optimal subset, referred to as a chunk combination, that fits within a given cost budget. Moreover, as shown in Fig.~\ref{fig:example}, the order of chunks in this combination critically affects RAG performance. Therefore, we incorporate the chunk combination order to denote it, and the objective is to determine the combination and ordering that yield the best benefit, formally defined as follows:
\begin{definition}[Optimal Chunk Combination Order Selection\label{def:occo}]
Let \(Q\) denote the input query, \(\{\chi_1, \chi_2, \ldots, \chi_k\}\) the set of candidate chunks,
\(\mathcal{B}\) the cost budget, and \(W(\Phi)\) the benefit of a chunk combination order
\(\Phi = \langle \chi_{\phi_1}, \ldots, \chi_{\phi_m} \rangle\), where \(\phi_i\) indicates the
position of each chunk in the sequence.
The objective is to identify the optimal chunk combination order \(\hat{\Phi}\) that maximizes
the overall benefit under the cost constraint:
\begin{equation}
\small
\hat{\Phi} = \arg\max_{\Phi} W(\Phi) \quad \text{s.t.} \quad \sum_{\chi_{i} \in \Phi} \text{cost}(\chi_i) \leq \mathcal{B}
\end{equation}
\vspace{-5mm}
\end{definition}

Importantly, solving the above problem is non-trivial because the benefit of a chunk combination cannot be obtained by simply summing the individual benefits of its constituent chunks. Instead, the overall benefit must be evaluated by invoking the reranker model on the whole combination (i.e., $W(\Phi)\neq \sum_{\chi_i\in \Phi} W(\chi_i)$).
To verify this, we conducted experiments to analyze the non-additivity of chunk benefits,
as detailed in our technical report~\cite{CARROT},
which demonstrates that all tested reranker models exhibit non-additive scoring behavior.
Furthermore, the problem is computationally challenging, as formally established by the following theorem.
\begin{theorem}
The Optimal Chunk Combination Order Selection problem is NP-hard.
\end{theorem}
The NP-hardness is proved via a polynomial-time reduction from the
\textit{Maximum Weighted Hyperclique Problem (MWHP)}~\cite{ausiello2012complexity},
which has been shown to be NP-hard, as discussed in the corresponding technical
report~\cite{CARROT}.

\section{Related Work\label{sec:related}}
\subsubsection{Retrieval-augmented Generation}

 RAG~\cite{chat2data,DBLP:journals/pvldb/tsg,DBLP:journals/pvldb/WangWKGXC24, naiverag} is widely used to handle knowledge-intensive NLP tasks. Current studies~\cite{DBLP:conf/acl/Mrag,DBLP:conf/iclr/selfrag,DBLP:conf/nips/rankrag,GraphRAG} focus on how to optimize RAG performance using different approaches, broadly classified as rank-based methods, graph-based methods, and tuning-based methods. In typical rank-based methods, a dense embedding-based retriever searches for relevant information from an external database, which is then used by the LLM generation process. To improve this pipeline, some studies~\cite{richrag,DBLP:conf/icde/LiYKWLX24} have focused on training retrievers to better suit the generation needs of LLMs, developing multi-step retrieval methods, and filtering out irrelevant information. Although there are many advanced retrievers~\cite{chat2data, RetClean, DBLP:journals/pvldb/WangWKGXC24,DBLP:conf/acl/hyde}, it is more promising to optimize the retriever and LLM together in an end-to-end process~\cite{NL2SQL,DBLP:journals/pvldb/ZeroEA}. For example, research~\cite{DBLP:conf/naacl/ShiMYS0LZY24,Completeness-Oriented} has focused on training retrievers and LLMs together, either simultaneously or in stages. For graph-based methods~\cite{GraphRAG, DBLP:conf/acl/JinXZRZL0TWM024, gretriever, raptor, DBLP:conf/nips/hipporag1}, studies try to use knowledge graphs to optimize external knowledge organization. However, these methods require frequent external LLM calls to generate the graph structure or guide actions, especially when the graph needs to be re-indexed or updated with new knowledge frequently, which incurs high computational costs. Therefore, tuning-based methods such as \cite{DBLP:conf/nips/rankrag,DBLP:conf/iclr/selfrag,DBLP:conf/naacl/tuning} use instruction tuning to improve response comprehensiveness and enable the LLM to self-reflect or rank knowledge without frequent re-indexing. However, these approaches often overlook the high computational cost in the tuning phase, which significantly impacts the practicality of RAG systems. To our knowledge, this paper is the first approach that considers the order of chunk combinations within computational cost constraints.

\subsubsection{Reranking for RAG}
Reranking methods~\cite{mulreranker,RL_reranker,DBLP:conf/acl/lightrerank, DBLP:conf/nips/proof} are crucial for improving retrieval performance in rank-based RAG methods, as they align retrieved chunks with LLM requirements and prioritize the content most conducive to generating high-quality responses. Traditional reranking approaches~\cite{DBLP:conf/acl/LiWLC18,DBLP:conf/aaai/Retriever-Reranker} typically rely on mid-sized language models, such as BERT or T5, to rank retrieved contexts. However, these models often struggle to capture semantic relationships between queries and contexts, especially in zero-shot or few-shot settings. Therefore, recent tuning-based research~\cite{DBLP:conf/nips/rankrag, DBLP:conf/iclr/selfrag, DBLP:conf/naacl/tuning} highlights the potential of instruction-tuned LLMs to improve context reranking by more accurately identifying relevant contexts, even in the presence of noise or irrelevant information. Despite these advancements, the full capacity of LLMs for reranking in RAG systems remains underutilized. In particular, studies have shown that the arrangement of chunks can affect LLM performance~\cite{DBLP:conf/acl/order}, emphasizing the need to consider the order of the combination of chunks in RAG tasks. However, existing methods are not well-suited for cases where optimal retrieval requires a specific chunk combination order, rather than isolated chunks. In this paper, we propose \textit{CARROT}, a cost-constrained retrieval optimization framework that addresses the overlooked importance of considering the order of chunk combinations.

\subsubsection{Reinforcement Learning for Large Language Models}  Recently, reinforcement learning (RL) has been increasingly utilized in various data management and RAG tasks. The RL technique can enable large language models to improve their generation ability by leveraging external knowledge sources, such as search engines~\cite{RL_reranker}. In particular, the human feedback~\cite{DBLP:conf/acl/Mrag,DBLP:conf/www/Mqueryrecomm} can be integrated to help models produce more accurate and contextually relevant responses through the RL framework. In addition, some query optimization approaches~\cite{chat2data,crafting,DBLP:journals/pvldb/diver} further refine retrieval processes, allowing model performance to inform query adjustments and ultimately enhance downstream task outcomes. In this work, we apply a lightweight RL technique MCTS to optimize the chunk combination order search process in the RAG system. We also introduce a configuration agent to guide the MCTS search process. To the best of our knowledge, this is the first approach to addressing this problem.

    \section{System Overview}

Existing RAG frameworks face three major challenges: capturing inter-chunk correlations,
handling the non-monotonic utility of chunk order, and adapting to diverse query domains.
To tackle these, we propose \textit{CARROT}, a learned system that retrieves the optimal chunk combination
under a given query and cost budget.
Its core component, the \textit{Optimal Chunk Combination Search} module, employs an MCTS-based
policy tree to sequentially explore chunk orders under cost constraints, effectively modeling
inter-chunk dependencies (\textbf{Challenge 1}) and non-monotonic utilities (\textbf{Challenge 2}).
Furthermore, the \textit{Configuration Agent} module automatically selects the best MCTS configuration
and reranker for different query domains, addressing \textbf{Challenge 3}.
We next overview these two modules.

\noindent
\textbf{1. Optimal Chunk Combination Search:} A straightforward approach to considering chunk correlations involves retrieving potential chunks from a vector database (Step 1 in Figure~\ref{fig:framework}) and exhaustively exploring all possible chunk combinations. However, this method incurs significant latency and computational costs. To mitigate this, we construct a policy tree (Step 2 in Figure~\ref{fig:framework}), reframing the optimal chunk combination search as a node search problem within the tree. Specifically, the root node of the policy tree represents an initial empty state, and each child node corresponds to a specific combination of chunks. We design an MCTS-based search algorithm to address this problem, which iteratively constructs
a policy tree for optimal chunk combination search. During the search process, both cost and
budget constraints are explicitly incorporated to guide decision-making.
Each iteration of the algorithm consists of four phases:
(1) \textit{Selection}: iteratively selecting child nodes from the root using a UCB-based utility function that balances exploration and exploitation while incorporating cost control;
(2) \textit{Expansion}: generating all feasible child nodes;
(3) \textit{Simulation}: performing parallel evaluations of the expanded nodes via reranker scoring; and
(4) \textit{Backpropagation}: updating path statistics with the obtained benefits.
In particular, we leverage parallel reranker evaluations to process multiple chunk combinations simultaneously, substantially improving computational efficiency.

\noindent \textbf{2. Configuration Inference:}
A naive approach to configuration tuning is to exhaustively evaluate all configurations and rerankers in parallel, which is prohibitively expensive for RAG systems. To address this, we introduce a \textit{Configuration Agent} that dynamically generates optimal settings—such as iteration count, cost coefficient, and exploration factor—based on the query domain and retrieved data.
Specifically, the configuration agent takes both the query and the retrieved chunks as input, fusing their embeddings through a data-aware mechanism to capture the correlation between the query intent and the actual data distribution, enabling configuration prediction conditioned on both query semantics and data characteristics. To
enhance the model's effectiveness, we employ a contrastive learning approach that uses positive and negative label pairs: positive labels correspond to query-data pairs with the same optimal reranker, while negative labels come from different optimal rerankers. A joint loss function is used to simultaneously optimize both the regression (for parameter tuning) and contrastive learning (to enhance label differentiation).

\noindent \textit{\textbf{RAG Pipeline.}}
Our framework pipeline is shown in Figure~\ref{fig:framework}.
We first generate an embedding for the input query, which is then used to retrieve potential chunks from the vector database. Both the query and retrieved chunks are fed into the configuration agent, which dynamically generates the optimal configuration by analyzing the query-data correlation through data-aware embedding fusion. Using the optimal configuration, we can search the policy tree to determine the optimal chunk combination and order from the retrieved potential chunks. Finally, this optimal chunk combination is used to construct the final prompt for the LLMs.

\begin{figure*}[h]
\centering
\vspace{-20pt}
\includegraphics[width=0.91\textwidth]{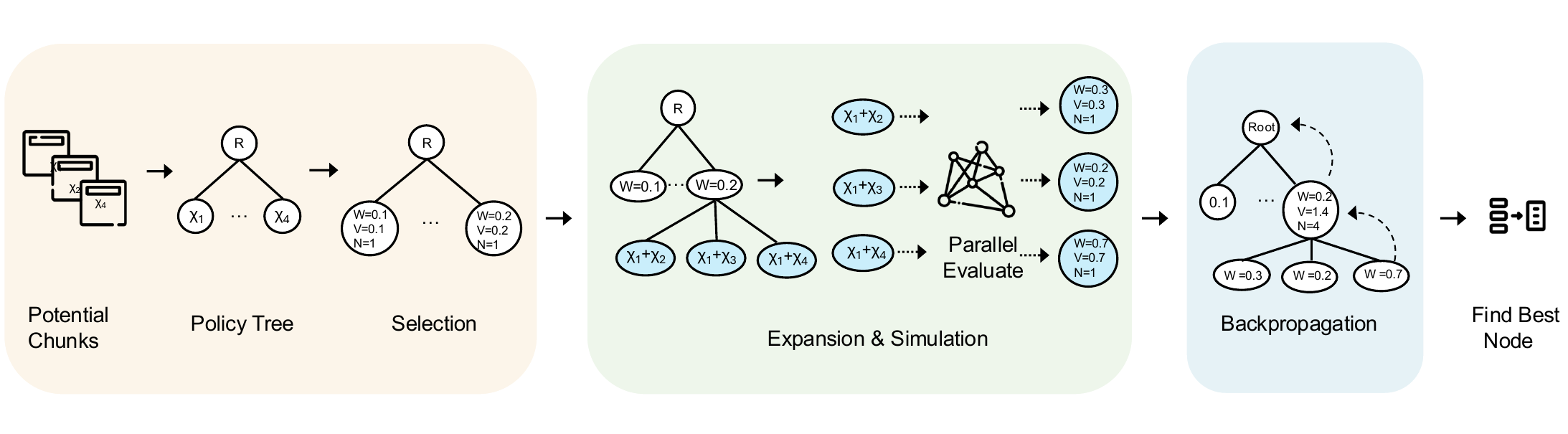}
\vspace{-15pt}
\caption{Workflow of MCTS-Based Policy Tree Search }
\vspace{-20pt}
\label{fig:MCTS}
\end{figure*}

    \section{Chunk Combination Search}

\noindent \textbf{Motivation.} The optimal chunk combination order selection problem presents two fundamental challenges. First, the search space grows exponentially: for $n$ candidate chunks, there exist $\sum_{k=1}^{n} P(n,k) = \sum_{k=1}^{n} \frac{n!}{(n-k)!}$ ordered combinations (e.g., $n=10$ yields over 9 million combinations), making exhaustive enumeration infeasible. Second, greedy strategies may fail due to the non-monotonic nature of chunk benefits. For example, while individual chunks may satisfy $W(\chi_1) > W(\chi_2)$, their chunk combination orders could yield $W(\langle \chi_2, \chi_3 \rangle) \gg W(\langle \chi_1, \chi_3 \rangle)$ in a real RAG scenario, as established in Section~\ref{sec:prelimi}. Once the greedy approach selects $\chi_1$ in the first step, it cannot backtrack to explore the superior combination $\langle \chi_2, \chi_3 \rangle$, thus missing the globally optimal chunk combination order. To address these challenges, we design a policy tree to organize the combination space and formulate the chunk combination order selection task as an optimal node search problem in the policy tree. We employ an MCTS-based search algorithm that strategically explores only a small subset of promising paths instead of enumerating all possibilities. The key insight is that MCTS iteratively expands and evaluates nodes along high-potential paths, accumulating node benefits through backpropagation, and ultimately returns the node with the highest reranker-assigned benefit among the explored paths. Unlike greedy selection, MCTS's exploration mechanism ensures that nodes with initially lower individual benefits can still be discovered if they lead to superior combinations.

\subsection{Policy Tree Search Modeling\label{sec:model}}

\noindent \textbf{Policy Tree.}
As illustrated in Figure~\ref{fig:MCTS}, we construct a policy tree to represent all possible orders of chunk combinations retrieved from the vector database. The root node corresponds to the initial state with no selected chunks, while each child node is generated by appending one additional chunk to the ordered sequence represented by its parent node.
Thus, each non-root node corresponds to an ordered combination of chunks, incrementally built through this process.
For example, if a node represents the combination \(\Phi_{v_1}=\langle\chi_1\rangle\), its child nodes may correspond to ordered extensions such as \(\Phi_{v_2}=\langle\chi_1, \chi_2\rangle\), \(\Phi_{v_3}=\langle\chi_1, \chi_3\rangle\), or \(\Phi_{v_4}=\langle\chi_1, \chi_4\rangle\).

\noindent \textit{Remark.} The entire policy tree does not need to be constructed and typically only a small portion of it is dynamically expanded during our MCTS-based search process.

\noindent \textbf{Optimal Node Search Modeling.} Each non-root node \(v_i\) in the policy tree $T$ represents a chunk combination order \(\Phi_{v_i}\), whose corresponding cost and benefit are denoted as \(\text{cost}(\Phi_{v_i})\) and \(W(\Phi_{v_i})\), respectively, as defined in Sec.~\ref{sec:prelimi}. Therefore, the task described in Definition~\ref{def:occo} can be reformulated as searching for the node within the cost budget \(\mathcal{B}\) that yields the highest benefit. Formally, this is expressed as:$\hat{v_i} = \arg\max_{v_i \subseteq T,~\text{cost}(\Phi_{v_i}) \leq \mathcal{B}} W(\Phi_{v_i})$.

\subsection{MCTS-Based Policy Tree Search\label{sec:search}}

Enumerating all possible nodes in the policy tree \(T\) is infeasible due to the exponential growth of the search space. To efficiently explore this space, we employ an MCTS-based approach that incrementally expands nodes to construct a partial tree \(T_{\text{MCTS}}\), from which a
solution can be derived without full enumeration.
In each iteration, the algorithm selects nodes to expand, evaluates their potential, and uses the evaluation results to guide the next round of search. To incorporate cost constraints and improve efficiency, we design a cost-aware node utility function to guide node selection and introduce a parallel evaluation mechanism to accelerate the scoring process.

\noindent \textbf{Cost-aware Node Utility \(\text{U}(v)\).} Because exhaustive exploration is infeasible, we guide the search with a UCB-style~\cite{DBLP:journals/jmlr/UCB} heuristic that balances exploitation and exploration while accounting for cost.
For each node \(v\), the algorithm maintains the cumulative reward \(V(v)\) (aggregated over historical visits) and the visit count \(N(v)\). We use \(N_{\text{parent}(v)}\) to denote the visit count of \(v\)'s parent node.
We define the cost-aware utility as
\begin{equation}
\label{for:utility}
\text{U}(v) \;=\; \frac{V(v)}{N(v)} \;+\; c\,\sqrt{\frac{\ln N_{\text{parent}(v)}}{N(v)}} \;-\; \lambda\,\frac{\text{cost}(\Phi_v)}{\mathcal{B}},
\end{equation}
where \(c\) and \(\lambda\) are tuning parameters. This extends standard UCB to our budget-constrained setting.

\noindent \textbf{Algorithm Overview}. As outlined in Algorithm~\ref{alg:1}, we initialize the root node of the policy tree and iteratively refine the tree until the iteration budget is exhausted. Each iteration performs four phases: (1) \textit{Selection}, traversing from root to an unexpanded node using node utility $\text{U}(v)$; (2) \textit{Expansion}, generating all possible child nodes within the cost budget; (3) \textit{Simulation}, evaluating all children in parallel and assigning them node benefits $W(\Phi_v)$; and (4) \textit{Backpropagation}, updating the cumulative reward \(V(\cdot)\) and visit count \(N(\cdot)\) along the path to the root using the obtained benefit \(W(\Phi_v)\). After completing all iterations, we obtain the MCTS policy tree $T_{MCTS}$ and select the node with the maximum reranker-assigned benefit \(W(\Phi_v)\) among all non-root nodes that satisfy the cost budget constraint. Notably, unlike traditional MCTS, which focuses on leaf nodes, our method considers nodes at all depths, as shorter combinations may outperform longer ones due to budget constraints or semantic coherence.

\begin{algorithm}[t]
\caption{MCTS-Based Policy Tree Search\label{alg:1}}
\small
\KwIn{$q$: a query; $D$: a set of candidate chunks; $\mathcal{B}$: total budget; $Reranker$: predicted reranker}
Initialize root node $v_0$ with query $q$ for the policy tree $T_{MCTS}$\;
\While{within iteration}{
    $v_{selected} = \text{Selection}(v_0, \mathcal{B})$\;
    $\{v_1, \ldots, v_k\} = \text{Expansion}(v_{selected}, D, \mathcal{B})$\;
    $\text{Simulation}(v_{selected}, \{v_1, \ldots, v_k\}, Reranker)$\;
    $\text{BackPropagation}(\{v_1, \ldots, v_k\})$\;
}
\Return{$\arg\max_{v \in T_{MCTS}, \text{cost}(\Phi_v) \leq \mathcal{B}} W(\Phi_v)$}
\end{algorithm}

\setcounter{algocf}{1}

\noindent \textbf{Detailed Phases.}
We now describe the five phases of the MCTS-based policy tree search.

\noindent \underline{\textit{(1) Selection (Function~\ref{alg:2}).}}  Starting from the root, the algorithm recursively selects child nodes guided by the utility function \(\text{U}(v)\) (Eq.~\ref{for:utility}), which balances exploitation, exploration, and cost. At each step, among available children, we select the one with the highest \(\text{U}(v)\) value. The exploration term $c\sqrt{\ln N_{\text{parent}(v)}/N(v)}$ in \(\text{U}(v)\) ensures that less-visited nodes receive priority, preventing premature convergence to locally optimal paths. Selection proceeds until reaching an unexpanded node, ensuring both deep and broad exploration of the search space.

\noindent \underline{\textit{(2) Expansion (Function~\ref{alg:3}).}}  Upon reaching an unexpanded node \(v\), all potential child nodes \(\{v_1, v_2, \ldots, v_k\}\) are generated by appending one additional chunk (each child cost does not exceed the budget $\mathcal{B}$) to the current combination. Unlike standard MCTS, which expands a single child per iteration, our approach expands all possible children simultaneously to prepare for parallel evaluation. In particular, we initialize each new node's cumulative reward and visit count to zero.

\noindent \underline{\textit{(3) Simulation: Parallel Evaluation (Function~\ref{alg:4}).}}
Instead of random rollouts, we evaluate all expanded nodes in parallel using the reranker $
[W(\Phi_{v_1}), \ldots, W(\Phi_{v_k})] = \text{Reranker}(Q, [\Phi_{v_1},\ldots, \Phi_{v_k}])$, where \(W(\Phi_{v_j})\) denotes the reranker-assigned benefit for combination \(\Phi_{v_j}\).
This design allows all \(k\) combinations to be scored in one model call (cost \(O(r)\))
rather than \(k\) separate calls (cost \(O(k \cdot r)\)),
leveraging transformer rerankers' batch processing for substantial efficiency gains.
For each \(v_j\), we record \(W(\Phi_{v_j})\) and compute its token cost \(\text{cost}(\Phi_{v_j})\).

\noindent \underline{\textit{(4) Backpropagation (Function~\ref{alg:5}).}}
Each child's benefit \(W(\Phi_{v_j})\) is propagated upward along the path to the root,
updating cumulative rewards \(V(\cdot)\) and visit counts \(N(\cdot)\).
Unlike traditional MCTS, which updates only the best path,
we backpropagate all children independently, preserving richer exploration statistics
to guide future iterations.

\noindent \underline{\textit{(5) Final Node Selection.}}
After all iterations, we perform a global search across explored nodes and select the one with the highest benefit under the cost budget: $\hat{v} = \arg\max_{v \in T_{MCTS},~\text{cost}(\Phi_v) \le \mathcal{B}} W(\Phi_{v})$. Unlike traditional MCTS methods, the final decision is based on the reranker-assigned benefit \(W(\Phi_v)\) rather than the averaged reward \(V(v)/N(v)\),
as \(W(\Phi_v)\) more accurately reflects the true relevance between query and chunk combination order. When multiple nodes have identical \(W(\Phi_v)\), we prefer nodes with more visits and greater depth.

\begin{figure}[t]
\begin{FunctionInline}
\caption{Selection\label{alg:2}}
\small
\SetKwFunction{FSelection}{Selection}
\KwIn{$v_i$: current tree node; $\mathcal{B}$: total budget}
\While{$v_i$.isFullyExpanded()}{
    \ForEach{child nodes $v_j$ of $v_i$}{
        compute utility $\text{U}(v_j) = \frac{V(v_j)}{N(v_j)} + c \sqrt{\frac{\ln N_{\text{parent}(v_j)}}{N(v_j)}} - \lambda \frac{\text{cost}(\Phi_{v_j})}{\mathcal{B}}$\;
    }
    $v_i=\arg\max_{v_j\in \text{Children}(v_i)} \text{U}(v_j)$\;
}
\Return{$v_i$}
\end{FunctionInline}
\vspace{-5pt}
\begin{FunctionInline}
\caption{Expansion\label{alg:3}}
\small
\SetKwFunction{FExpansion}{Expansion}
\KwIn{$v$: unexpanded node to expand; $D$: set of candidate chunks; $\mathcal{B}$: total budget}
\tcp{Create all possible child nodes}
Expand \(v\) by using the candidate chunks in \(D\) to construct child nodes \(\{v_1, v_2, \ldots, v_k\}\) whose costs do not exceed the budget \(\mathcal{B}\)\;
\ForEach{child nodes $v_j$}{
    calculate and record $\text{cost}(\Phi_{v_j})$\;
    initialize $N(v_j)=0, V(v_j)=0$\;
}
\Return{$\{v_1, v_2, \ldots, v_k\}$}
\end{FunctionInline}
\vspace{-5pt}
\begin{FunctionInline}
\caption{Simulation\label{alg:4}}
\small
\SetKwFunction{Simulation}{Simulation}
\KwIn{$v$: parent node; $\{v_1, \ldots, v_k\}$: expanded children; $Reranker$: predicted reranker}
\tcp{Parallel evaluation via reranker}
compute benefit $[W(\Phi_{v_1}), \ldots, W(\Phi_{v_k})] = \text{Reranker}(Q, [\Phi_{v_1},\ldots, \Phi_{v_k}])$\;
\ForEach{child nodes $v_j$}{
    record $W(\Phi_{v_j})$\;
}
\end{FunctionInline}
\vspace{-5pt}
\begin{FunctionInline}
\caption{BackPropagation\label{alg:5}}
\small
\SetKwFunction{FBackProp}{BackPropagation}
\KwIn{$\{v_1, v_2, \ldots, v_k\}$: the expanded child nodes}
\tcp{Backpropagate each child}
\ForEach{child nodes $v_j$}{
    $v_{current} = v_j$\;
    \Repeat{$v_{current}$ is None}{
        $V(v_{current}) = V(v_{current}) + W(\Phi_{v_j})$\;
        $N(v_{current}) = N(v_{current}) + 1$\;
        $v_{current} = \text{Parent}(v_{current})$\;
    }
}
\end{FunctionInline}
\end{figure}

\begin{figure*}[!t]
  \centering
  \includegraphics[width=0.88\textwidth]{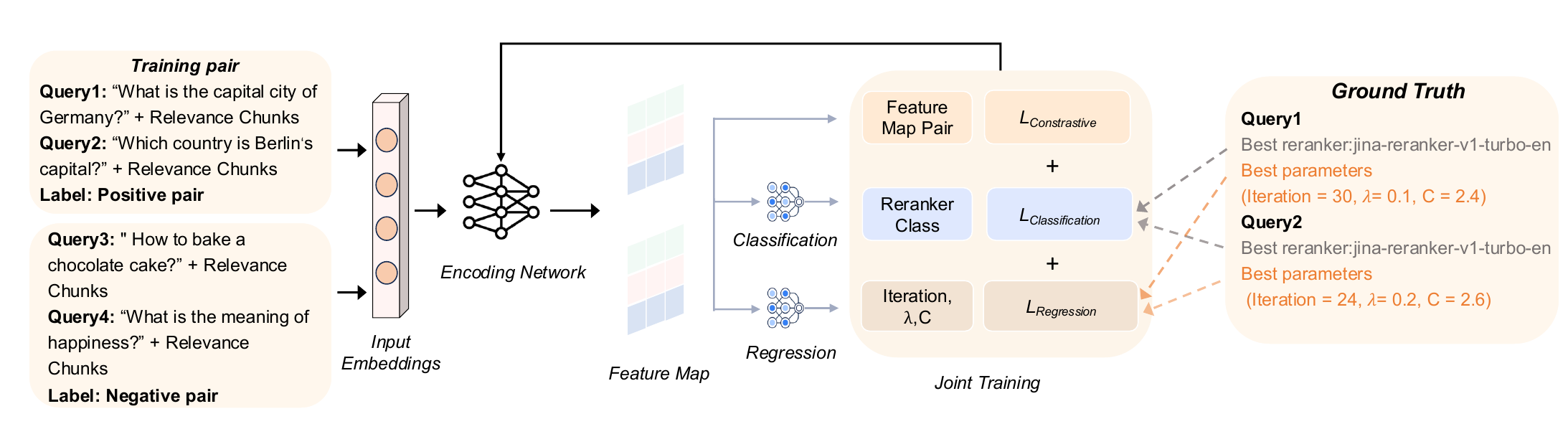}
  \vspace{-10pt}
  \caption{Overview of Configuration Agent}
  \label{fig:cons}
  \vspace{-10pt}
\end{figure*}

\begin{figure}[t]
  \centering
  \includegraphics[width=0.8\columnwidth]{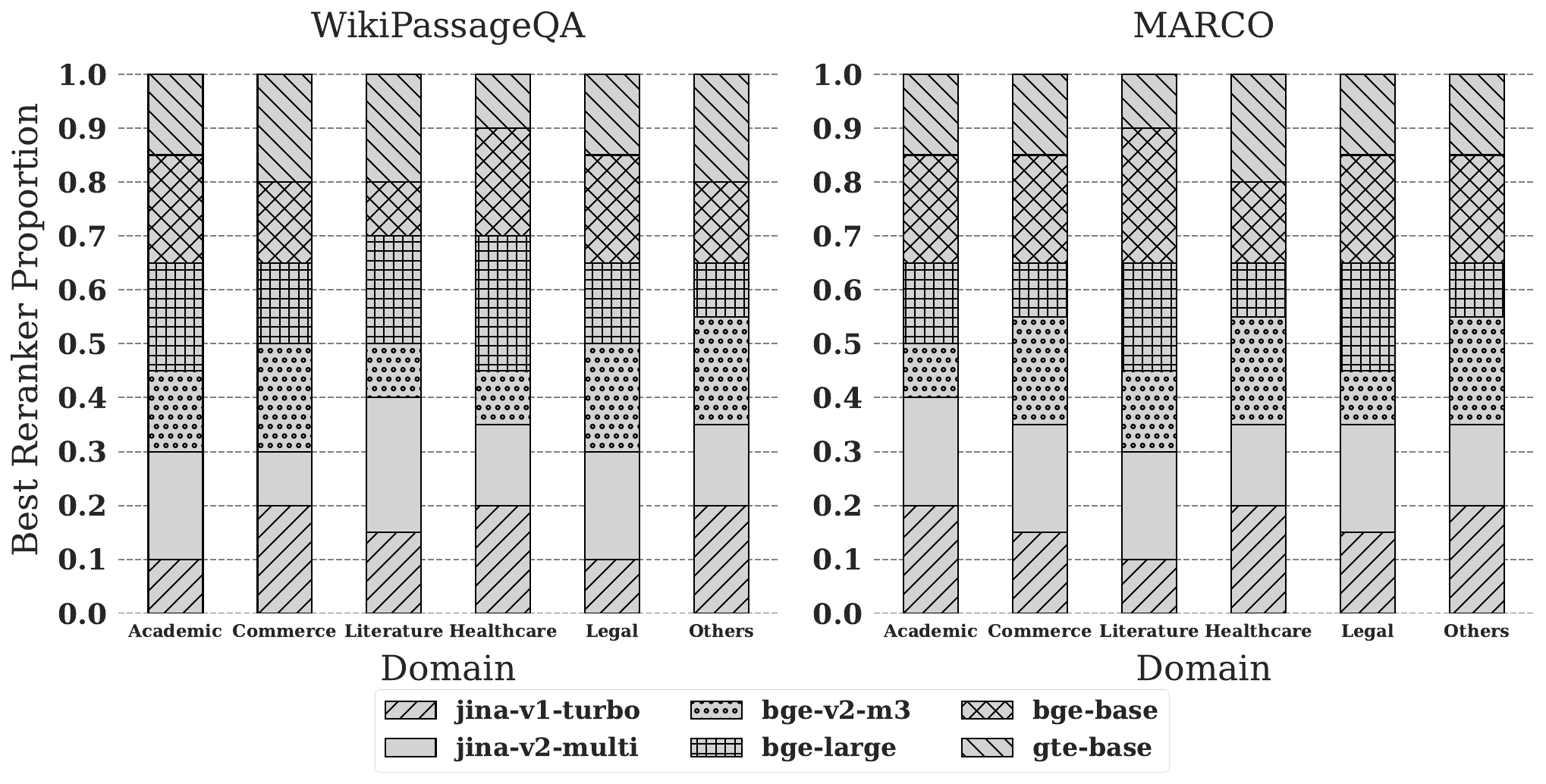}
  \caption{Best Reranker Distribution}
  \label{fig:Distribution}
  \vspace{-10pt}
\end{figure}

\noindent \textbf{Complexity Analysis.}
Let $N$ be the number of candidate chunks, $\mathcal{B}$ the token budget, $s$ the fixed chunk token cost, and $r$ the maximum cost of a single batched reranker call. Any budget-feasible combination contains at most $D_{\mathcal{B}}=\min\{N,\lfloor \mathcal{B}/s\rfloor\}$ chunks.\\
\noindent \underline{\textit{(1) Time Complexity}}: Each of the $I$ iterations performs selection, expansion, simulation, and backpropagation. (i) Selection traverses $O(D_{\mathcal{B}})$ levels with $O(N)$ children per level, yielding $O(D_{\mathcal{B}} \cdot N)$ complexity; (ii) Expansion creates up to $N$ children, contributing $O(N)$; (iii) Simulation evaluates all expanded children in a single batched reranker call, costing $O(r)$; (iv) Backpropagation updates $O(N)$ children across $O(D_{\mathcal{B}})$ levels, giving $O(D_{\mathcal{B}} \cdot N)$.
Thus, the total time complexity is $O(I \cdot (D_{\mathcal{B}} \cdot N + r))$. A smaller $\mathcal{B}$ reduces $D_{\mathcal{B}}$, shrinking the search space. In practice, $I$ is predicted by the configuration agent. As shown in Fig.~\ref{fig:candidates_different}, $I$ remains a small value, ensuring efficient search.\\
\noindent \underline{\textit{(2) Space Complexity}}: Space is allocated to store the nodes of the tree. In the $i$-th iteration, we expand $k_i$ nodes, where $k_i \leq N$. Over $I$ iterations, this results in a total of $\sum_{i=1}^I k_i\leq I\cdot N$ nodes, leading to a space complexity of $O(I \cdot N)$.

    \section{Configuration Agent}

\noindent \textbf{Motivation.} The MCTS process requires critical configurations including reranker selection, iteration count, exploration coefficient, and cost coefficient. Our preliminary experiments (details in our technical report~\cite{CARROT}) reveal that optimal configurations vary across queries and retrieved data. For instance, Fig.~\ref{fig:Distribution} shows that different query domains (Academic, Commerce, etc.) exhibit distinct best-performing rerankers, validating the necessity of adaptive configuration. Therefore, we design a configuration agent that predicts optimal MCTS configurations based on both query and chunk characteristics. A natural alternative is to directly predict chunk combination orders via a learned model. However, this is impractical: (i) the $N!$ combinatorial space makes supervised data collection infeasible; (ii) non-monotonic utility functions where $W(A \cup B) \neq W(A) + W(B)$ are difficult for neural networks to approximate; (iii) predicting chunk combination orders requires structured outputs from the $N!$ space, whereas predicting MCTS hyperparameters reduces to lightweight regression. Accordingly, we present the agent framework in Section~\ref{sec:agent_framework} and the learning pipeline in Section~\ref{sec:joint}.

\vspace{-0.1in}
\subsection{Model Framework\label{sec:agent_framework}}
\noindent \textbf{Design Rationale.} A straightforward approach is to use a multilayer perceptron (MLP) classifier to predict the optimal reranker for each query. However, due to the high semantic and domain diversity of RAG queries, such classifiers often perform poorly, as shown in our experimental result in Table~\ref{tab:carrot_ablation}.
To overcome this limitation, we adopt a Siamese network trained with contrastive learning, which encourages similar query representations to be closer while pushing apart dissimilar ones, promoting cross-domain generalization~\cite{DBLP:conf/nips/Contrastive}.

Figure~\ref{fig:cons} illustrates the overall architecture, which comprises two main modules for transforming the input into feature maps.
First, the embedding module generates representations for input queries and their retrieved chunks.
Then, the encoding network refines these embeddings into feature maps, which are used by classification or regression layers to produce the final configuration outputs.
The following sections describe each component in detail.

\noindent \underline{\textit{(1) Input Embedding.}} Given a query and its retrieved chunks from the vector database, we employ the BGE-M3 model~\cite{DBLP:journals/corr/bgem3} to generate 1024-dimensional embeddings for the query and the mean-pooled representation of retrieved chunks. Encoding the retrieved chunks enables the agent to capture query-related data characteristics, enhancing its sensitivity to contextual nuances. These embeddings preserve semantic information, facilitating efficient comparison and classification by the encoding network. This design improves retrieval relevance across diverse query types by explicitly modeling the relationship between query intent and data distribution.

\noindent \underline{\textit{(2) Feature Map Generation with Encoding Network.\label{sec:contrastive_learning}}}
To optimize reranker selection across diverse query types, we employ an encoding network that jointly learns classification and configuration representations. A Siamese network with three fully connected layers fuses query and chunk embeddings, capturing query–data correlations for context-dependent predictions. Both branches share weights with linear layers and ReLU activations, reducing the concatenated 2048-dimensional input ($x \oplus c$) through layers of 512, 256, and 128 dimensions. The output layer produces (i) a classification result identifying the optimal reranker and (ii) regression outputs predicting MCTS configurations (iterations, exploration coefficient $c$, and cost coefficient $\lambda$). By conditioning on both query and chunk embeddings, this architecture adaptively controls various search parameters. For example, by predicting iterations, it adjusts search depth: fewer iterations for simple query-data pairs where lightweight search suffices, more for complex ones requiring deeper exploration.
\vspace{-0.1in}

\subsection{Joint Training}
\label{sec:joint}
In this section, we present the training pipeline for the configuration agent (Figure~\ref{fig:cons}). It integrates three joint tasks: classification for reranker selection, regression for MCTS hyperparameter prediction, and contrastive learning to further enhance representation quality.
\subsubsection{Classification and regression losses}
Let the query and its relevant chunk embeddings be denoted as $x$ and $c$, respectively, where $\oplus$ denotes the concatenation operation. The encoding network is represented by $f_{\theta}$, with a classification head $g_{\theta}$ and a regression head $h_{\theta}$ in the output layer.
The classification loss is defined as $L_{\text{cla}}(\theta) = F_{\text{cla}}\big(g_{\theta}(f_{\theta}(x \oplus c)),\, y_{\text{true}}\big)$, where $y_{\text{true}}$ is the ground-truth optimal reranker obtained through actual reranker execution, and $F_{\text{cla}}$ denotes the cross-entropy loss between predicted and true labels.
Similarly, the regression loss is given by  $L_{\text{reg}}(\theta) = F_{\text{reg}}\big(h_{\theta}(f_{\theta}(x \oplus c)),\, p_{\text{true}}\big)$, where $p_{\text{true}}$ represents the optimal MCTS hyperparameters (iterations, exploration coefficient $c$, and cost coefficient $\lambda$) identified via grid search on real query executions, and $F_{\text{reg}}$ denotes the mean squared error (MSE) loss.

\subsubsection{Contrastive Learning} To efficiently distinguish between different query domains and recommend the optimal configuration for each query, we utilize contrastive learning to bring queries with the same optimal reranker closer together while pushing apart embeddings from different reranker classes. \\
\noindent \textbf{Contrastive pairs preparation.}
First, we obtain ground truth labels via grid search over parameters using ROUGE-L, evaluating different parameter combinations to select configurations yielding optimal response quality. Next, we generate contrastive pairs based on these optimal annotations: positive pairs share the same optimal reranker (promoting minimal embedding distance), and negative pairs use different rerankers (maximizing embedding distance). We construct approximately 10,000 training pairs. The labeling process only requires pairwise preference comparisons between configurations, and we deploy Llama3-8B locally to generate training pairs. Since the contrastive pairs are constructed from query-chunk embeddings with ground truth labels, the agent learns query-context patterns that share similar optimal configurations. These patterns become domain-invariant representations, enabling generalization to different downstream LLMs and unseen domains without retraining.

\noindent \textbf{Contrastive loss.}
As illustrated in Figure~\ref{fig:cons}, for a given pair $(x_i \oplus c_i,x_j \oplus c_j)$ with similarity indicator $I_{ij}$ ($I_{ij}=0$ for positive pairs with the same reranker, $I_{ij}=1$ for negative pairs with different rerankers), we first generate their corresponding feature maps with the encoding model $f_{\theta}$. These feature maps are then utilized to compute the contrastive loss \(L_{\text{con}}\). As shown in Algorithm~\ref{alg:joint}, the contrastive loss uses the similarity indicator $I_{ij}$ to handle both positive and negative pairs in a unified formulation. This loss function is designed to ensure that query-data pairs with the same reranker are brought closer together in the embedding space, while those with different rerankers are distanced.

\begin{algorithm}[t]
\caption{Joint Contrastive Training}\label{alg:joint}
\small
\KwIn{Query embedding pairs $(x_i,x_j)$, chunk embeddings $(c_i,c_j)$, labels $(y_i,y_j)$, params $(p_i,p_j)$, similarity indicator $I_{ij}$ (0 for same class, 1 for different class)}
\KwOut{models $f_\theta$, $g_\theta$ and $h_\theta$}
Initialize $f_\theta$ with projection head\;
\For{epoch $=1$ to $N$}{
    \For{batch $(x_i,x_j,c_i,c_j,I_{ij},y_i,y_j,p_i,p_j)$}{
        $z_i\gets f_\theta(x_i \oplus c_i)$, $z_j\gets f_\theta(x_j \oplus c_j)$
        $y_i^{pred}\gets g_\theta(z_i), p_i^{\text{pred}} \gets h_\theta(z_i)$\;
        $y_j^{pred}\gets g_\theta(z_j), p_j^{\text{pred}} \gets h_\theta(z_j)$\;
        $L_{\text{con}} \gets (1-I_{ij})\|z_i-z_j\|^2 + I_{ij}\max(0,\gamma-\|z_i-z_j\|)^2$
        $L_{\text{cla}} \gets CE(y_i^{pred},y_i) + CE(y_j^{pred},y_j)$\tcp*{CE is the cross-entropy loss}
        $L_{\text{reg}} \gets \|p_i^{\text{pred}}-p_i\|^2 + \|p_j^{\text{pred}}-p_j\|^2$\;
        Update $\theta$ using $\nabla_\theta(L_{\text{con}} + L_{\text{cla}} + L_{\text{reg}})$\;
    }
}
\end{algorithm}
\subsubsection{Whole training process}
As shown in Algorithm \ref{alg:joint}, we implement the complete training process by designing a loss function $L_{\text{total}}$ that combines contrastive, classification, and regression losses as follows: $L_{\text{total}}(\theta) = L_{\text{con}}(\theta) + L_{\text{cla}}(\theta) + L_{\text{reg}}(\theta)$.
In particular, the contrastive loss $L_{\text{con}}(\theta)$ encourages the embeddings of queries with the same optimal reranker to be close together, while pushing apart the embeddings of queries with different rerankers. The classification loss $L_{\text{cla}}(\theta)$ aids the model in correctly identifying the reranker using cross-entropy, and the regression loss $L_{\text{reg}}(\theta)$ minimizes the error in predicting the optimal MCTS configuration.

\noindent \textbf{Complexity Analysis} We analyze the computational complexity of both training and inference procedures for the Configuration Agent. (1) During offline training, we adopt a contrastive learning framework that processes \( N \) training pairs in each batch. The computational cost of generating embeddings is \( O(N \cdot T_E) \), where \( T_E \) denotes the embedding generation time for a single sample. Subsequently, each embedding is passed through our lightweight configuration network, incurring an additional cost of \( O(N \cdot T_S) \), where \( T_S \) represents the processing time per sample in this smaller network. Since the configuration network is significantly smaller and faster than the embedding network (e.g., BGE-M3), we have \( T_S \ll T_E \). Thus, the total training time complexity is dominated by embedding generation: \( O(N \cdot T_E + N \cdot T_S) = O(N \cdot T_E) \). This implies that the training complexity scales linearly with the number of training samples. (2) For online inference, suppose the query workload size is \( |Q| \) and the network latency per query is \( C \). The overall inference complexity is then \( O(C \cdot |Q|) \). In our empirical study in Table~\ref{tab:carrot_ablation}, both classification prediction and parameter prediction process queries within an acceptable efficiency range.

    \vspace{-0.1in}
\section{Experiment}

The experimental study intends to answer the following questions:
\begin{itemize}[leftmargin=10.2pt]
\setlength{\itemsep}{0pt}
\setlength{\parsep}{0pt}
\setlength{\parskip}{0pt}

    \item \textbf{RQ1}:
    (1) How effective is our
    \textit{CARROT} for the cost-constrained RAG pipeline compared to other methods? (2) Does \textit{CARROT} consistently maintain its effectiveness across different LLMs?

       \item \textbf{RQ2}: How efficient is \textit{CARROT} compared to different baselines?

    \item \textbf{RQ3}: How does \textit{CARROT} perform with large-scale datasets when evaluating its scalability?

    \item \textbf{RQ4}: How does the performance of \textit{CARROT} vary across different cost budget constraints?

    \item \textbf{RQ5}: What is the effectiveness of each component in \textit{CARROT}?

\end{itemize}

\begin{table}[!t]
\small
  \caption{Statistics of datasets used in the experiments.}
  \label{table:dataset}
  \centering
  \setlength{\tabcolsep}{2pt}
   \renewcommand{\arraystretch}{0.9}
  \begin{tabular}{ccccc}
    \toprule
    \textbf{Dataset} & \textbf{\#train} & \textbf{\#dev} & \textbf{\#test} & \textbf{\#p} \\
    \midrule
    MARCO & 808,731 & 101,093 & 101,092 & 8,841,823 \\
    WikiPassageQA & 3,332 & 417 & 416 & 244,136 \\
    HotpotQA & 90,564 & 7,405 & 14,810 & 5,000,000+ \\
    \bottomrule
  \end{tabular}
  \vspace{-0.25in}
\end{table}
\vspace{-0.1in}
\subsection{Experimental Setting}

\noindent \textbf{Environment.} We integrate our framework with the RAG framework LlamaIndex~\cite{llamaindex} and Faiss~\cite{douze2024faiss}. Experiments used dual Intel Xeon Gold 6326 CPUs (32 cores, 64 threads), 512\,GB DDR4 RAM, and 960\,GB NVMe SSD. The configuration agent is implemented in PyTorch 2.4, with data collection and training performed on 8$\times$A5000 GPUs. We choose OpenRouter~\cite{openrouter} to deploy our inference service.

\noindent \textbf{Datasets.} To evaluate the performance of \textit{CARROT} across diverse scenarios, we conduct experiments on three datasets with differing task focuses (Table \ref{table:dataset}): (1) \textit{WikiPassageQA}~\cite{wikipassage} is a question-answering benchmark containing 4,165 questions and over 100,000 text chunks, designed for non-factoid question answering and passage retrieval. (2) \textit{MARCO}~\cite{DBLP:conf/nips/marco}  is a comprehensive dataset tailored for natural language processing tasks, primarily emphasizing question answering and retrieval. (3) \textit{HotpotQA}~\cite{hotpotqa} is a question-answering benchmark designed for multi-hop reasoning tasks, consisting of approximately 113K QA pairs sourced from Wikipedia.

\begin{table*}[t]
\centering
\vspace{-0.15in}
\caption{Effectiveness Comparison and Token Cost Across Datasets (All methods are built upon Llama3-8B. Best in \textbf{bold}, second-best \underline{underlined}). On.=Average online token cost (per-query).}
\vspace{-0.05in}
\footnotesize
\setlength{\tabcolsep}{1.5pt}
\scalebox{0.82}{
\begin{tabular}{lccccccccccccccccccccccccc}
\toprule
& \multicolumn{10}{c}{WikiPassageQA} & \multicolumn{10}{c}{MARCO} & \multicolumn{5}{c}{HotpotQA} \\
\cmidrule(lr){2-11} \cmidrule(lr){12-21} \cmidrule(lr){22-26}
\textbf{Method} & \multicolumn{3}{c}{\textbf{256}} & \multicolumn{3}{c}{\textbf{512}} & \multicolumn{3}{c}{\textbf{1024}} & \textbf{On.}$\downarrow$ & \multicolumn{3}{c}{\textbf{256}} & \multicolumn{3}{c}{\textbf{512}} & \multicolumn{3}{c}{\textbf{1024}} & \textbf{On.}$\downarrow$ & \textbf{256} & \textbf{512} & \textbf{1024} & \textbf{On.}$\downarrow$ \\
\cmidrule(lr){2-4} \cmidrule(lr){5-7} \cmidrule(lr){8-10} \cmidrule(lr){11-11} \cmidrule(lr){12-14} \cmidrule(lr){15-17} \cmidrule(lr){18-20} \cmidrule(lr){21-21} \cmidrule(lr){22-22} \cmidrule(lr){23-23} \cmidrule(lr){24-24} \cmidrule(lr){25-26}
& \textbf{R1} & \textbf{RL} & \textbf{B-1} & \textbf{R1} & \textbf{RL} & \textbf{B-1} & \textbf{R1} & \textbf{RL} & \textbf{B-1} & & \textbf{R1} & \textbf{RL} & \textbf{B-1} & \textbf{R1} & \textbf{RL} & \textbf{B-1} & \textbf{R1} & \textbf{RL} & \textbf{B-1} & & \textbf{F1} & \textbf{F1} & \textbf{F1} & \\
\midrule
RAPTOR & 0.328 & 0.306 & 0.345 & 0.322 & 0.301 & 0.333 & 0.335 & 0.305 & 0.370 & 1024 & 0.386 & 0.356 & 0.371 & 0.393 & 0.366 & 0.370 & 0.338 & 0.316 & 0.328 & 1024 & 0.248 & 0.252 & 0.245 & 1024 \\
NaiveRAG & 0.327 & 0.302 & 0.333 & 0.321 & 0.297 & 0.321 & 0.334 & 0.303 & 0.338 & 1024 & 0.398 & 0.369 & 0.385 & 0.395 & 0.368 & 0.372 & 0.337 & 0.312 & 0.324 & 1024 & 0.243 & 0.248 & 0.240 & 1024 \\
HYDE & 0.311 & 0.301 & 0.323 & 0.306 & 0.286 & 0.311 & 0.318 & 0.290 & 0.328 & 1024 & 0.367 & 0.338 & 0.352 & 0.373 & 0.348 & 0.349 & 0.321 & 0.300 & 0.312 & 1024 & 0.228 & 0.235 & 0.225 & 1024 \\
ColBERT & 0.329 & 0.304 & 0.335 & 0.324 & 0.300 & 0.326 & 0.333 & 0.307 & 0.336 & 1024 & 0.376 & 0.357 & 0.358 & 0.375 & 0.349 & 0.351 & 0.324 & 0.309 & 0.319 & 1024 & 0.247 & 0.253 & 0.244 & 1024 \\
Graph-CoT & 0.362 & 0.343 & 0.324 & 0.359 & 0.341 & 0.321 & 0.364 & 0.345 & 0.326 & 14237 & 0.364 & 0.345 & 0.349 & 0.362 & 0.343 & 0.347 & 0.360 & 0.342 & 0.345 & 14576 & 0.294 & 0.297 & 0.295 & 14011 \\
G-Retriever & 0.356 & 0.337 & 0.318 & 0.353 & 0.335 & 0.315 & 0.358 & 0.339 & 0.320 & 13239 & 0.358 & 0.339 & 0.343 & 0.356 & 0.337 & 0.341 & 0.354 & 0.336 & 0.339 & 12428 & 0.281 & 0.284 & 0.282 & 13003 \\
FLARE & 0.367 & 0.348 & 0.330 & 0.364 & 0.346 & 0.327 & 0.370 & 0.351 & 0.333 & 16913 & 0.370 & 0.351 & 0.355 & 0.368 & 0.349 & 0.353 & 0.366 & 0.348 & 0.352 & 16584 & 0.305 & 0.307 & 0.305 & 15022 \\
GraphRAG & 0.337 & 0.315 & 0.311 & 0.322 & 0.302 & 0.291 & 0.331 & 0.290 & 0.304 & 1024 & 0.376 & 0.343 & 0.357 & 0.370 & 0.342 & 0.351 & 0.307 & 0.295 & 0.304 & 1004 & 0.298 & 0.318 & 0.291 & 1001 \\
HippoRAG2 & 0.363 & 0.345 & 0.328 & 0.360 & 0.343 & 0.325 & 0.367 & 0.348 & 0.331 & 1024 & 0.367 & 0.348 & 0.352 & 0.365 & 0.347 & 0.350 & 0.363 & 0.346 & 0.349 & 1003 & 0.311 & 0.327 & 0.313 & 1028 \\
\midrule
\textbf{CARROT upper} & \textbf{0.437} & \textbf{0.416} & \textbf{0.395} & \textbf{0.426} & \textbf{0.406} & \textbf{0.423} & \textbf{0.444} & \textbf{0.423} & \textbf{0.420} & 923 & \textbf{0.435} & \textbf{0.414} & \textbf{0.431} & \textbf{0.425} & \textbf{0.397} & \textbf{0.411} & \textbf{0.447} & \textbf{0.426} & \textbf{0.420} & 822 & \textbf{0.347} & \textbf{0.357} & \textbf{0.339} & 1011 \\
\textbf{CARROT w/o Agent} & 0.390 & 0.364 & 0.337 & 0.372 & 0.347 & 0.379 & 0.388 & 0.362 & 0.368 & \underline{838} & 0.401 & 0.374 & 0.389 & 0.393 & 0.372 & 0.377 & 0.393 & 0.364 & 0.368 & \textbf{725} & 0.299 & 0.312 & 0.307 & 1021 \\
\textbf{CARROT} & \underline{0.432} & \underline{0.401} & \underline{0.392} & \underline{0.413} & \underline{0.381} & \underline{0.394} & \underline{0.419} & \underline{0.386} & \underline{0.397} & \textbf{811} & \underline{0.425} & \underline{0.391} & \underline{0.413} & \underline{0.419} & \underline{0.386} & \underline{0.400} & \underline{0.423} & \underline{0.393} & \underline{0.405} & \underline{789} & \underline{0.320} & \underline{0.349} & \underline{0.336} & \textbf{1007} \\
\bottomrule
\end{tabular}
}
\vspace{-0.1in}
\label{tab:rouge_comparison}
\end{table*}

\begin{table*}[t]
\centering
\caption{Effectiveness Comparison Across Different LLMs (best in \textbf{bold}, second-best \underline{underlined})}
\vspace{-0.05in}
\renewcommand{\arraystretch}{0.9}
\small
\setlength{\tabcolsep}{3pt}

\resizebox{0.95\textwidth}{!}{
\begin{tabular}{c|c|cccccccccc|ccc}
\hline
\textbf{Dataset} & \textbf{LLM} & RAPTOR & NaiveRAG & HYDE & ColBERT & Graph-CoT & G-Retriever & FLARE & GraphRAG & HippoRAG2 & \textbf{CARROT upper} & \textbf{CARROT w/o Agent} & \textbf{CARROT} \\
\hline

\multirow{3}{*}{\parbox[c]{3cm}{\centering WikiPassageQA \\ (Rouge-1 score $\uparrow$)}}
& Llama3-8B & 0.328 & 0.327 & 0.311 & 0.329 & 0.362 & 0.356 & 0.367 & 0.337 & 0.363 & \textbf{0.437} & 0.390 & \underline{0.432} \\
& DeepseekV3 & 0.348 & 0.347 & 0.330 & 0.348 & 0.371 & 0.368 & 0.389 & 0.358 & 0.369 & \textbf{0.463} & 0.413 & \underline{0.459} \\
& GPT-4o & 0.341 & 0.340 & 0.324 & 0.344 & 0.367 & 0.371 & 0.382 & 0.351 & 0.361 & \textbf{0.454} & 0.406 & \underline{0.449} \\

\hline

\multirow{3}{*}{\parbox[c]{3cm}{\centering MARCO \\ (Rouge-1 score $\uparrow$)}}
& Llama3-8B & 0.386 & 0.398 & 0.367 & 0.376 & 0.364 & 0.358 & 0.370 & 0.376 & 0.367 & \textbf{0.435} & 0.401 & \underline{0.425} \\
& DeepseekV3 & 0.409 & 0.422 & 0.389 & 0.379 & 0.419 & 0.428 & 0.429 & 0.399 & 0.410 & \textbf{0.461} & 0.425 & \underline{0.454} \\
& GPT-4o & 0.401 & 0.414 & 0.382 & 0.372 & 0.411 & 0.418 & 0.422 & 0.391 & 0.402 & \textbf{0.452} & 0.417 & \underline{0.440} \\
\hline

\multirow{3}{*}{\parbox[c]{3cm}{\centering HotpotQA \\ (F1 score $\uparrow$)}}
& Llama3-8B & 0.248 & 0.243 & 0.228 & 0.247 & 0.294 & 0.281 & 0.305 & 0.298 & 0.311 & \textbf{0.347} & 0.299 & \underline{0.320} \\
& DeepseekV3 & 0.263 & 0.258 & 0.242 & 0.262 & 0.311 & 0.298 & 0.319 & 0.316 & 0.329 & \textbf{0.368} & 0.317 & \underline{0.341} \\
& GPT-4o & 0.259 & 0.254 & 0.238 & 0.256 & 0.305 & 0.291 & 0.311 & 0.307 & 0.318 & \textbf{0.355} & 0.309 & \underline{0.331} \\
\hline
\end{tabular}}

\vspace{-0.15in}
\label{tab:llm_dataset_comparison}
\end{table*}

\noindent \textbf{Baselines.} We compare \textit{CARROT} against representative rank-based RAG methods, including \textit{RAPTOR}~\cite{raptor}, \textit{NaiveRAG}~\cite{naiverag}, \textit{HYDE}~\cite{DBLP:conf/acl/hyde}, and \textit{FLARE}~\cite{DBLP:conf/emnlp/FLARE}, to demonstrate improvements in retrieval relevance and response quality. We also evaluate graph-based approaches such as \textit{GraphRAG}~\cite{GraphRAG}, \textit{Graph-CoT}~\cite{DBLP:conf/acl/JinXZRZL0TWM024}, and \textit{G-Retriever}~\cite{DBLP:conf/nips/Gretriever}, highlighting our efficiency advantages. Finally, we include comparisons with tuning-based methods, despite their substantial computational overhead and infrastructure requirements. For instance, \textit{ColBERT}~\cite{DBLP:conf/sigir/colbert} requires 3 hours to index 9M passages on four V100 GPUs, while \textit{RankRAG}~\cite{DBLP:conf/nips/rankrag} and \textit{Self-RAG}~\cite{DBLP:conf/iclr/selfrag} demand 128 A100 GPUs for 16 hours and 4 A100 GPUs for 12 hours, respectively. We therefore include \textit{ColBERT} as a representative baseline for fairness, noting that such tuning-based methods are often impractical for large-scale deployment due to their prohibitive training and indexing costs.
Details of all baselines are provided in our technical report~\cite{CARROT}.

In addition, we remove the configuration agent from our method as a baseline to evaluate its impact on the performance of \textit{CARROT}, referring to this version as \textit{CARROT w/o Agent}. Finally, we implement a method called \textit{CARROT Upper} to establish an upper bound by enumerating all possible chunk combinations within the top-5 chunk candidates and selecting the optimal chunk combination order. We adopt Llama3-8B~\cite{DBLP:journals/corr/llama} as the default large language model for these methods.

\noindent \textbf{Hyper-parameter Settings.} The hyper-parameters for \textit{CARROT} are automatically determined by the configuration agent, while \textit{NaiveRAG} does not require hyper-parameters. For other baseline methods, we ensure consistency by using identical hyper-parameters for fair comparisons. Specifically, we set the exploration coefficient to 2.4, the number of iterations to 10, and the cost coefficient \(\lambda\) to 0.1. Due to the computational constraints imposed by combinatorial complexity, we set the candidate chunks to 5 in the \textit{CARROT Upper} case. By default, we set the fixed chunk size to 256 and the budget to 1024 for fair ablation studies. For offline data collection, we sample 1K seed queries from three datasets and perform grid search over 6 rerankers with 20 parameter combinations per query.

\noindent \textbf{Learning Setting.} In our method, the configuration agent is trained using contrastive learning. The hyper-parameters used during this process include a margin for contrastive loss (margin=1.0), learning rate (lr=0.001), batch size (32), number of epochs (num\_epochs=60), and the embedding model (i.e., BAAI/bge-m3~\cite{DBLP:journals/corr/bgem3}).

\noindent \textbf{Evaluation Metrics.} Following \cite{DBLP:conf/nips/rankrag,richrag,DBLP:conf/acl/Mrag}, we utilize ROUGE-1, ROUGE-L \cite{lin-2004-rouge}, and BLEU-1~\cite{DBLP:conf/wmt/BLEU} for single-hop retrieval tasks, and F1 score~\cite{powers2020evaluation} for multi-hop retrieval scenarios. To evaluate efficiency, we measure the retrieval latency across different methods.

\subsection{Performance Comparison }

\subsubsection{\textbf{RQ1-(1)}: Effectiveness Comparison Across Datasets and Chunk Sizes}

As shown in Table~\ref{tab:rouge_comparison}, we evaluate the effectiveness of \textit{CARROT} compared to various baselines on three datasets: WikiPassageQA, MARCO, and HotpotQA. The first two are single-hop datasets, while the last is a multi-hop dataset. For methods that support token-level budget constraints (e.g., \textit{RAPTOR}, \textit{NaiveRAG}, and \textit{HYDE}), we fix the token budget to 1024. For other methods, we report results without budget limitations. To ensure fairness, all methods are built upon Llama3-8B. The experimental results lead to the following observations:

\noindent (1) \textit{CARROT} consistently outperforms all baselines across different datasets. Specifically, it achieves about 30\% relative gain on both single-hop and multi-hop tasks. As expected, it does not surpass the \textit{CARROT upper}, which represents a nearly exhaustive search over chunk combinations.

\noindent (2) Optimal chunk sizes vary across methods and datasets. For instance, \textit{CARROT} performs best with a chunk size of 256 in the single-hop datasets and 512 in the multi-hop setting. Meanwhile, the \textit{CARROT upper} achieves its best results with a chunk size of 1024 in the single-hop dataset while 512 in multi-hop dataset. Notably, the average number of tokens used in HotpotQA is higher than in single-hop datasets, likely due to the increased complexity of multi-hop reasoning, which benefits from larger chunk sizes that capture more context.

\noindent (3) \textit{CARROT upper} achieves better performance at a modest increase in cost. For example, on WikiPassageQA, the R1 score improves from 0.432 to 0.437 while the average token usage increases from 811 to 923, demonstrating that exhaustive search with larger search spaces provides diminishing returns, validating the cost-quality trade-off of \textit{CARROT}.

\subsubsection{\textbf{RQ1-(2):} Effectiveness Comparison Across LLMs}
As shown in Table~\ref{tab:llm_dataset_comparison}, we evaluated \textit{CARROT} across three LLMs: Llama3-8B~\cite{DBLP:journals/corr/llama}, DeepseekV3~\cite{DBLP:journals/corr/deepseekv3}, and GPT-4o~\cite{GPT}. The evaluation utilizes Rouge-1 for WikiPassageQA, MARCO and F1 scores for HotpotQA, with a standardized cost constraint of $1024$ tokens. The results indicate that \textit{CARROT} consistently outperforms baseline methods across all LLMs, with particularly strong performance when integrated with DeepseekV3, highlighting our method's generalizability.

\subsubsection{\textbf{RQ2:} Efficiency Evaluation}
\begin{figure}[!t]
  \vspace{-0.05in}
    \centering
    \begin{subfigure}[b]{0.23\textwidth}
        \centering
        \includegraphics[width=\textwidth]{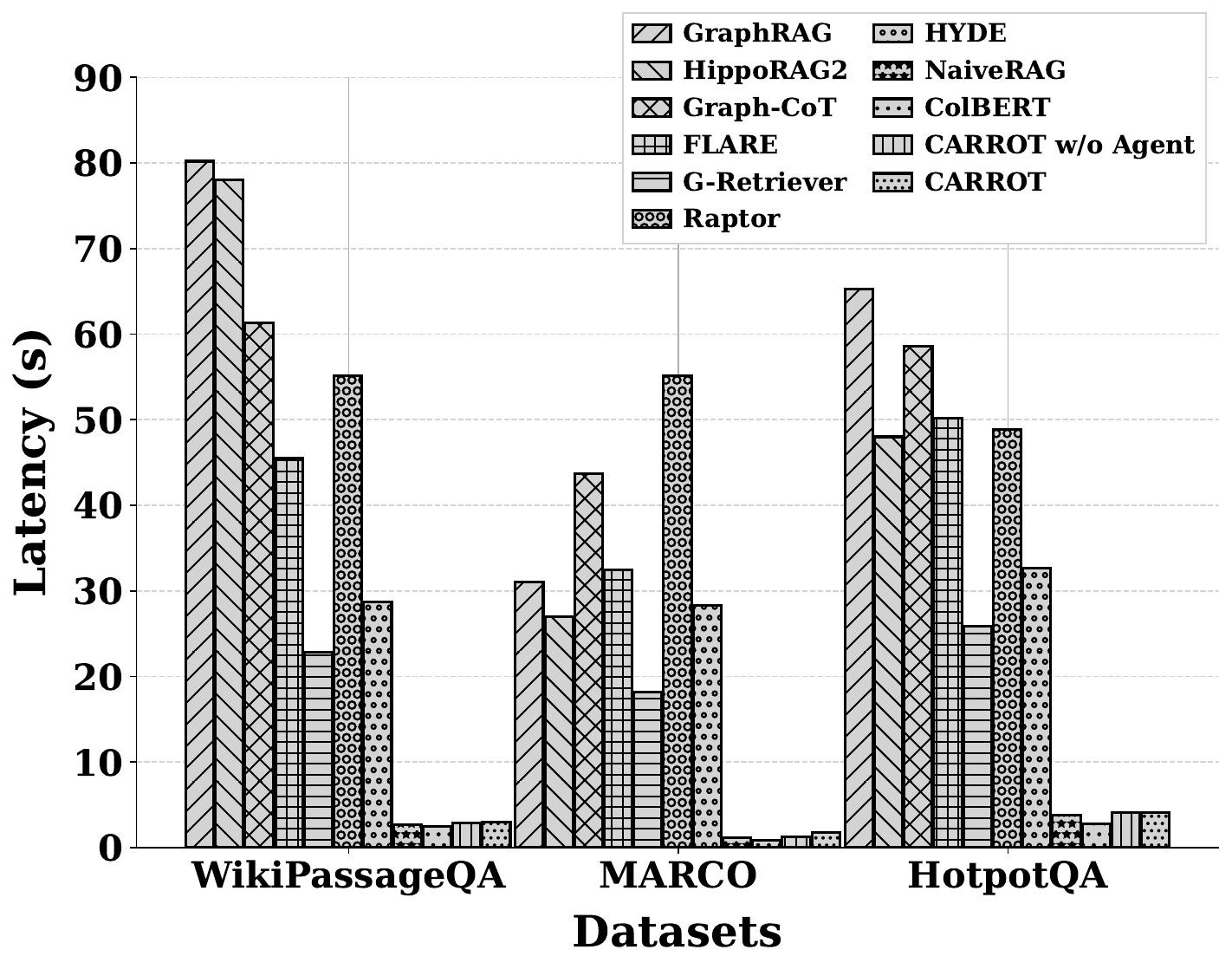}
        \vspace{-0.1in}
        \caption{Efficiency Comparison Across Datasets}
        \label{fig:efficiency_comparison}
    \end{subfigure}
    \hfill
    \begin{subfigure}[b]{0.23\textwidth}
        \centering
        \includegraphics[width=\textwidth]{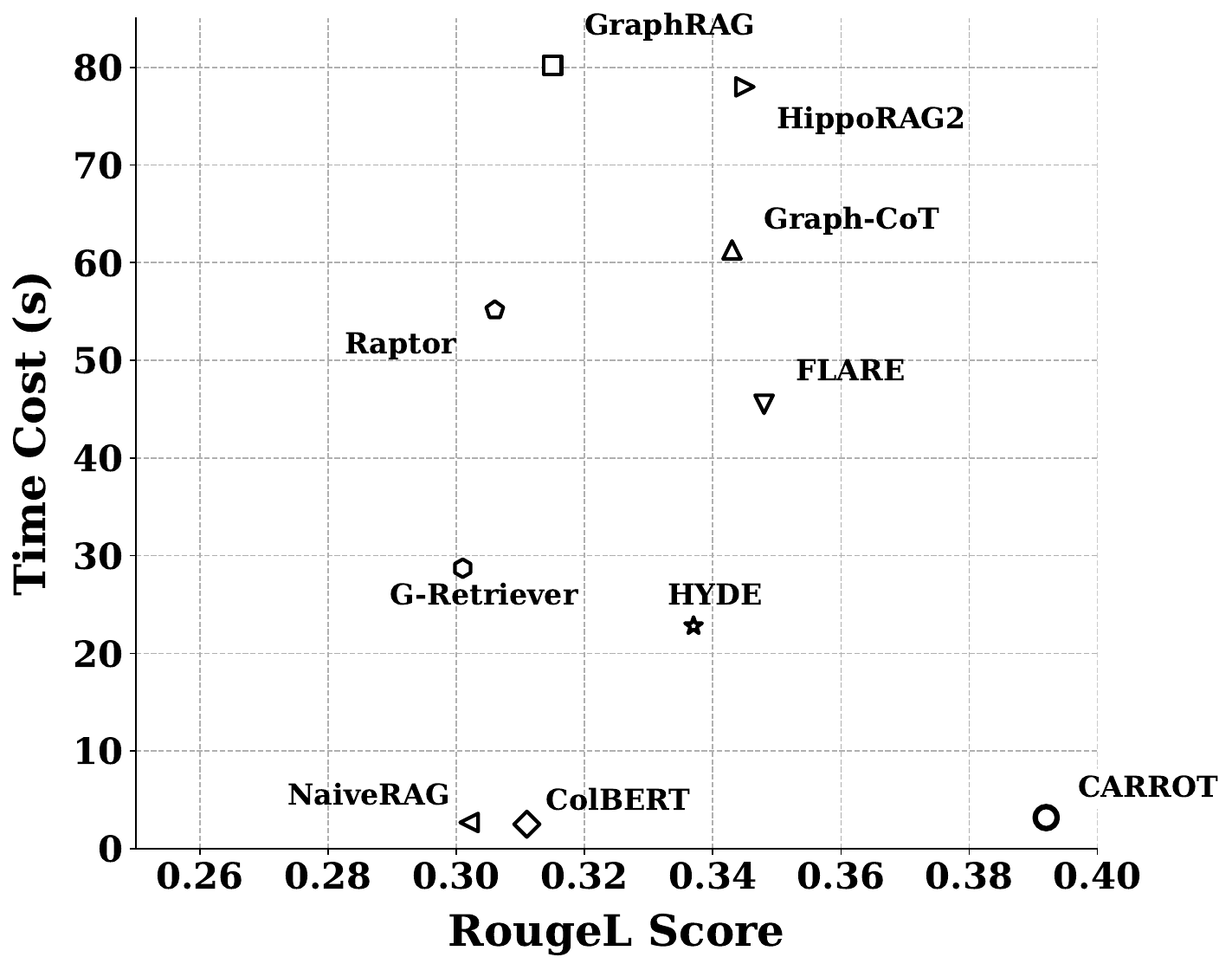}
        \caption{Time Cost vs Response Quality Tradeoff}
        \label{fig:quality_tradeoff}
    \end{subfigure}
    \caption{Efficiency and Quality Trade-off Analysis}
    \label{fig:efficiency_quality_analysis}
      \vspace{-0.3in}
\end{figure}

As depicted in Figure~\ref{fig:efficiency_comparison}, we evaluate the retrieval efficiency by comparing the latency between different methods across the three datasets, with the chunk size set at 256. The results show that \textit{NaiveRAG} requires the least time for retrieval as it directly invokes the reranker model and uses the vector index to calculate similarities. On the other hand, \textit{CARROT} requires more time as it employs a configuration agent to predict optimal parameters and utilizes a policy tree to search for the optimal chunk combination. Conversely, \textit{CARROT w/o Agent}, which operates with default parameter settings, demonstrates reduced retrieval time compared to \textit{CARROT}. Additionally, other baseline methods, which involve invoking LLMs to process retrieval, exhibit higher latency. For instance, \textit{RAPTOR} and Graph-based methods, which use an external LLM for summarization or extraction, showed significantly reduced efficiency.

Figure~\ref{fig:quality_tradeoff} demonstrates that \textit{CARROT} achieves high response quality within acceptable time costs. The comparable latencies between \textit{NaiveRAG} and \textit{CARROT} result from key optimizations: (1) Minimal LLM calls: \textit{CARROT} invokes the LLM only once at final inference after tree search completion. (2) Shallow search depth: optimal gold passages are typically covered by several chunks, resulting in shallow tree depths. (3) Parallel expansion: parallel mechanisms enable concurrent computation of multiple candidate nodes' utility scores without significantly increasing external model calls.

\begin{table}[!t]
\centering
\caption{\small{Cost Breakdown on WikiPassageQA (Llama3-8B)}.}
\label{tab:training_comparsion}
\footnotesize
\setlength{\tabcolsep}{2.5pt}
\renewcommand{\arraystretch}{0.92}
\begin{tabular}{l|cc|cc|cc}
\hline
\multirow{2}{*}{\textbf{Method}} & \multicolumn{2}{c|}{\textbf{Offline Indexing}} & \multicolumn{2}{c|}{\textbf{Online}} & \multicolumn{2}{c}{\textbf{Quality}} \\
\cline{2-7}
 & \raisebox{0.5ex}{\textbf{Time(s)}} & \raisebox{0.5ex}{\textbf{Token}} & \raisebox{0.5ex}{\textbf{Time(s)}} & \raisebox{0.5ex}{\textbf{Token}} & \raisebox{0.5ex}{\textbf{R1}$\uparrow$} & \raisebox{0.5ex}{\textbf{RL}$\uparrow$} \\
\hline
NaiveRAG & 61.2 & -- & 2.3 & 1024 & 0.327 & 0.302 \\
HYDE & 66.1 & -- & 28.7 & 1024 & 0.311 & 0.301 \\
ColBERT & 1731 & -- & 2.5 & 1024 & 0.329 & 0.304 \\
G-Retriever & 1281.1 & -- & 22.8 & 13239 & 0.356 & 0.337 \\
FLARE & 70.0 & -- & 45.5 & 16913 & 0.367 & 0.348 \\
\hline
\hline
RAPTOR & 21301.0 & 0.9M & 55.1 & 1024 & 0.328 & 0.306 \\
Graph-CoT & 28888.1 & 2.1M & 61.3 & 14237 & 0.362 & 0.343 \\
GraphRAG & 55003.2 & 2.3M & 80.2 & 1024 & 0.337 & 0.315 \\
HippoRAG2 & 32401.3 & 1.7M & 78.0 & 1024 & 0.363 & 0.345 \\
\hline
\textbf{CARROT} & \textbf{66.9} & -- & \textbf{3.1} & \textbf{811} & \textbf{0.432} & \textbf{0.401} \\
\hline
\end{tabular}
\vspace{-0.25in}
\end{table}

\begin{table*}[!t]
    \centering
    \vspace{-0.2in}
    \caption{Effectiveness comparison vs. budgets (best in \textbf{bold}, second-best \underline{underlined}). On.=Online token cost (per-query).}
      \vspace{-0.05in}
    \label{tab:budget}
    \renewcommand{\arraystretch}{0.9}
    \scalebox{0.8}{
    \begin{tabular}{l| cccccc |cccccc| cccccc}
        \hline
        \multirow{3}{*}{\textbf{Methods}} & \multicolumn{6}{c}{\textbf{Budget=1024}} & \multicolumn{6}{c}{\textbf{Budget=2048}} & \multicolumn{6}{c}{\textbf{Budget=8192}} \\
        \cline{2-19}
        & \multicolumn{2}{c}{\textbf{WikiPassageQA}} & \multicolumn{2}{c}{\textbf{MARCO}} & \multicolumn{2}{c}{\textbf{HotpotQA}} & \multicolumn{2}{c}{\textbf{WikiPassageQA}} & \multicolumn{2}{c}{\textbf{MARCO}} & \multicolumn{2}{c}{\textbf{HotpotQA}} & \multicolumn{2}{c}{\textbf{WikiPassageQA}} & \multicolumn{2}{c}{\textbf{MARCO}} & \multicolumn{2}{c}{\textbf{HotpotQA}} \\
        \cline{2-19}

        & \textbf{R1$\uparrow$} & \textbf{On.}$\downarrow$ & \textbf{R1$\uparrow$} & \textbf{On.}$\downarrow$ & \textbf{F1$\uparrow$} & \textbf{On.}$\downarrow$ & \textbf{R1$\uparrow$} & \textbf{On.}$\downarrow$ & \textbf{R1$\uparrow$} & \textbf{On.}$\downarrow$ & \textbf{F1$\uparrow$} & \textbf{On.}$\downarrow$ & \textbf{R1$\uparrow$} & \textbf{On.}$\downarrow$ & \textbf{R1$\uparrow$} & \textbf{On.}$\downarrow$ & \textbf{F1$\uparrow$} & \textbf{On.}$\downarrow$ \\
        \hline
        RAPTOR & 0.328 & 1024 & 0.386 & 1024 & 0.248 & 1024 & 0.334 & 2048 & 0.389 & 2048 & 0.255 & 2048 & 0.330 & 8192 & 0.381 & 8192 & 0.247 & 8192 \\
        NaiveRAG & 0.327 & 1024 & 0.398 & 1024 & 0.243 & 1024 & 0.333 & 2048 & 0.401 & 2048 & 0.250 & 2048 & 0.328 & 8192 & 0.392 & 8192 & 0.242 & 8192 \\
        HYDE & 0.311 & 1024 & 0.367 & 1024 & 0.228 & 1024 & 0.317 & 2048 & 0.371 & 2048 & 0.236 & 2048 & 0.312 & 8192 & 0.363 & 8192 & 0.227 & 8192 \\
                ColBERT & 0.329 & 1024 & 0.376 & 1024 & 0.247 & 1024 & 0.335 & 2048 & 0.379 & 2048 & 0.254 & 2048 & 0.330 & 8192 & 0.371 & 8192 & 0.246 & 8192 \\

        Graph-CoT & 0.362 & 14156 & 0.364 & 14089 & 0.262 & 14823 & 0.365 & 14298 & 0.367 & 14312 & 0.269 & 15067 & 0.361 & 14421 & 0.363 & 14487 & 0.260 & 15213 \\
        G-Retriever & 0.356 & 13102 & 0.358 & 13187 & 0.270 & 13356 & 0.359 & 13245 & 0.361 & 13312 & 0.276 & 13489 & 0.354 & 13378 & 0.356 & 13456 & 0.268 & 13612 \\
        FLARE & 0.367 & 16789 & 0.370 & 16102 & 0.301 & 16867 & 0.370 & 16912 & 0.373 & 16278 & 0.309 & 17023 & 0.366 & 17045 & 0.369 & 16421 & 0.303 & 17189 \\
        GraphRAG & 0.337 & 1024 & 0.376 & 1004 & 0.298 & 1001 & 0.343 & 2048 & 0.381 & 2048 & 0.305 & 2048 & 0.335 & 8192 & 0.369 & 8192 & 0.292 & 8192 \\
        HippoRAG2 & 0.363 & 1024 & 0.367 & 1003 & 0.311 & 1028 & 0.369 & 2048 & 0.372 & 2048 & 0.318 & 2048 & 0.362 & 8192 & 0.365 & 8192 & 0.309 & 8192 \\
        \hline
        \textbf{CARROT w/o Agent} & 0.390 & \underline{823} & 0.401 & \underline{841} & 0.299 & \underline{987} & 0.372 & \underline{856} & 0.393 & \textbf{712} & 0.312 & 1018 & 0.388 & \underline{879} & 0.393 & \textbf{748} & 0.307 & 1035 \\
        \textbf{CARROT upper} & \textbf{0.437} & 912 & \textbf{0.435} & 938 & \textbf{0.347} & 997 & \textbf{0.426} & 934 & \textbf{0.425} & 867 & \textbf{0.357} & \underline{1012} & \textbf{0.444} & 971 & \textbf{0.447} & 835 & \textbf{0.339} & \underline{1027} \\
        \textbf{CARROT} & \underline{0.432} & \textbf{798} & \underline{0.425} & \textbf{776} & \underline{0.320} & \textbf{991} & \underline{0.417} & \textbf{823} & \underline{0.419} & \underline{801} & \underline{0.349} & \textbf{1008} & \underline{0.420} & \textbf{847} & \underline{0.420} & \underline{819} & \underline{0.336} & \textbf{1015} \\
        \hline
    \end{tabular}}
    \vspace{-0.10in}
\end{table*}

\begin{table}
\centering
\caption{Ablation Study of Components (WikiPassageQA)}
\vspace{-0.05in}
\footnotesize
\setlength{\tabcolsep}{3pt}
\renewcommand{\arraystretch}{0.9}
\scalebox{0.9}{
\begin{tabular}{@{}lrr@{\hskip 8pt}rr@{}}
\toprule
\multirow{2}{*}{\textbf{Methods} (Chunk Size=256)} & \multicolumn{2}{c}{\textbf{Quality}} & \multicolumn{2}{c}{\textbf{Online Cost}} \\
\cmidrule(lr){2-3}\cmidrule(lr){4-5}
& \textbf{R1}$\uparrow$ & \textbf{RL}$\uparrow$ & \textbf{Tok.} & \textbf{Time} \\
\midrule
\textbf{Top-k+JinaV2} & 0.327 & 0.302 & 1024 & 2.3 \\
\textbf{w Agent-Only} & 0.297 & 0.271 & 795 & 2.5 \\
\textbf{w Beam Search} & 0.380 & 0.354 & 820 & 3.0 \\
\textbf{w Bandit} & 0.362 & 0.337 & 815 & 2.9 \\
\textbf{w MLP} & 0.341 & 0.318 & 992 & 3.5 \\
\textbf{w/o Agent} & 0.390 & 0.364 & 838 & 2.6 \\
\textbf{w/o Predicted Reranker} & 0.367 & 0.342 & 805 & 2.7 \\
\textbf{w/o Predicted C} & 0.394 & 0.367 & 815 & 2.8 \\
\textbf{w/o Predicted $\lambda$} & 0.391 & 0.364 & 812 & 2.8 \\
\textbf{w/o Parallel Expansion} & 0.401 & 0.374 & 840 & 16.1 \\
\midrule
\textbf{CARROT} & \textbf{0.432} & \textbf{0.401} & \textbf{811} & \textbf{3.1}\rlap{$^\dagger$} \\
\bottomrule
\end{tabular}
}
\par\vspace{1pt}
{\scriptsize\raggedright $^\dagger$\textbf{Latency breakdown}: Embed 0.2s (6.5\%), Retrieval 0.5s (16.1\%), Config Agent 0.5s (16.1\%), Simulation 1.5s (48.4\%), Tree Ops 0.3s (9.7\%), Other 0.1s (3.2\%).\par}
\vspace{-0.2in}
\label{tab:carrot_ablation}
\end{table}

Table~\ref{tab:training_comparsion} decomposes computational costs into two dimensions: Offline Indexing and Online retrieval. Offline Indexing captures per-corpus preprocessing costs that must be repeated for each new corpus, e.g., GraphRAG requires 2.3M tokens for WikiPassageQA indexing. \textit{CARROT} requires negligible per-corpus indexing of 66.9s. For online retrieval, \textit{CARROT} achieves 811 tokens per query with R1=0.432, demonstrating superior token efficiency compared to baselines, making \textit{CARROT} suitable for large-scale deployments.

\begin{figure*}[!t]
  \centering
  \hspace{-0.2in}
  \begin{minipage}[t]{0.55\textwidth}
    \centering
    \begin{minipage}[t]{0.5\textwidth}
      \centering
      \includegraphics[width=\textwidth]{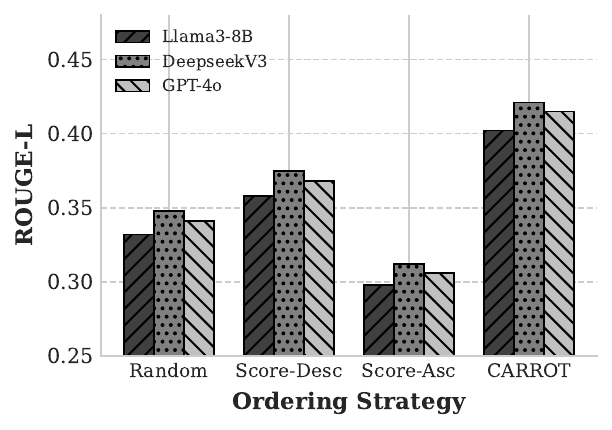}
      \vspace{-0.1in}
      \centerline{\small (a) Across LLMs}
      \label{fig:order_sensitivity_models}
    \end{minipage}
    \hfill
    \hspace{-0.2in}
    \begin{minipage}[t]{0.5\textwidth}
      \centering
      \includegraphics[width=\textwidth]{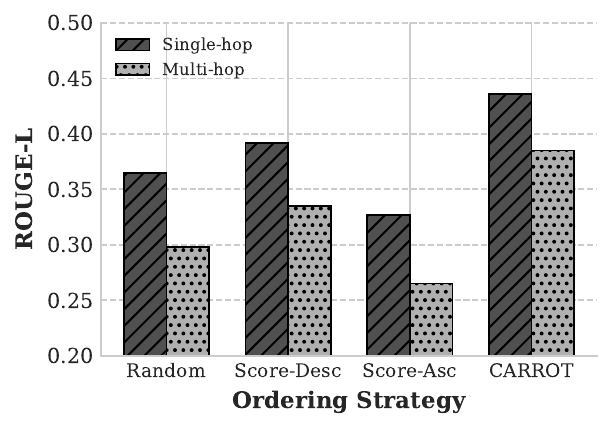}
      \vspace{-0.1in}
      \centerline{\small (b) Across query types}
      \label{fig:order_sensitivity_querytypes}
    \end{minipage}
    \vspace{-0.1in}
    \captionof{figure}{Chunk order sensitivity analysis}
    \label{fig:order_sensitivity}
  \end{minipage}
  \hspace{-0.3in}
  \hfill
  \begin{minipage}[t]{0.45\textwidth}
    \centering
    \begin{minipage}[t]{0.5\textwidth}
      \centering
      \includegraphics[width=\textwidth]{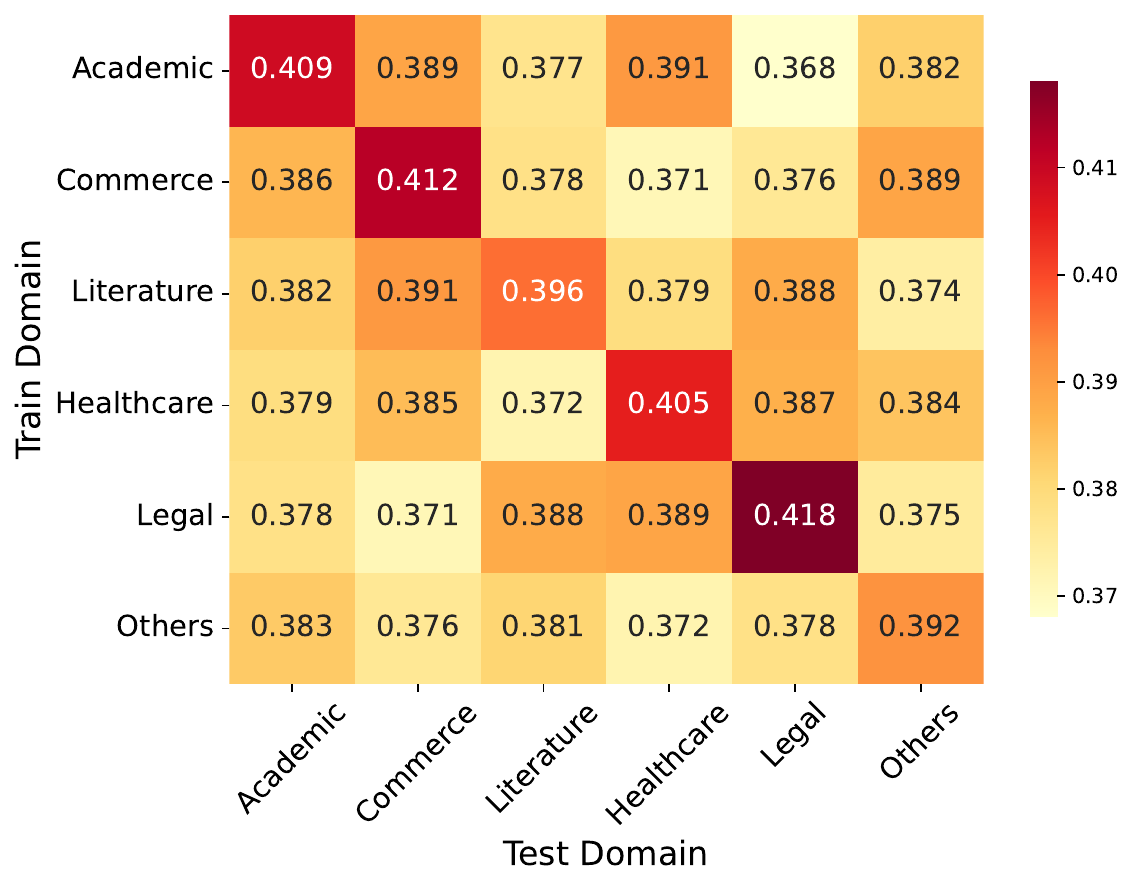}
      \vspace{-0.1in}
      \centerline{\small (a) Domain Generalization}
      \label{fig:domain_generalization}
    \end{minipage}
    \hfill
    \hspace{-0.2in}
    \begin{minipage}[t]{0.5\textwidth}
      \centering
      \includegraphics[width=\textwidth]{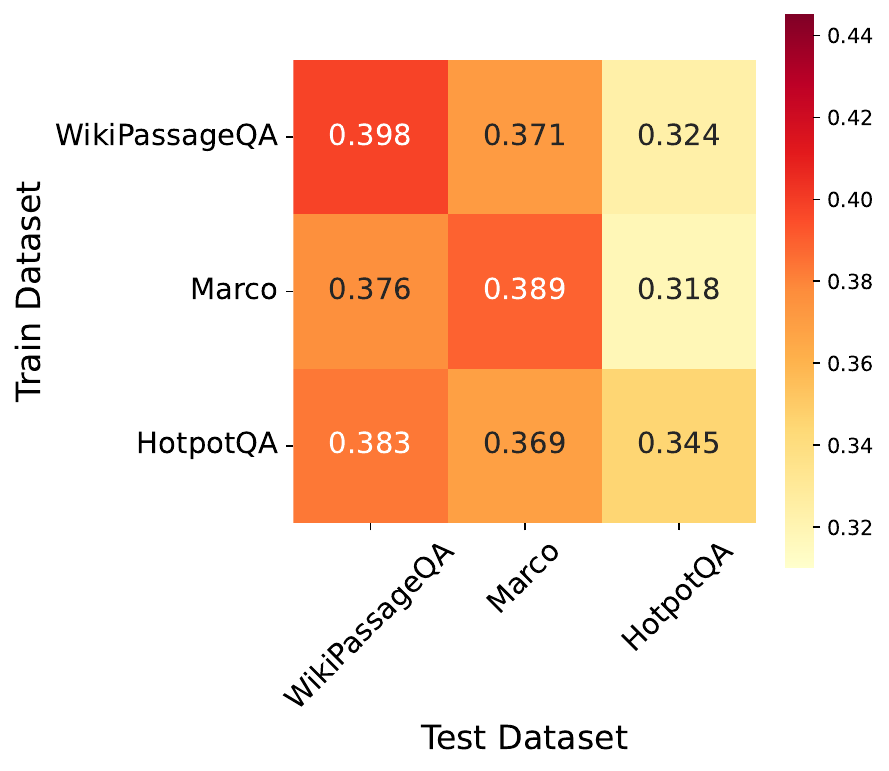}
      \vspace{-0.1in}
      \centerline{\small (b) Data Generalization}
      \label{fig:data_generalization}
    \end{minipage}
    \vspace{-0.1in}
    \captionof{figure}{Data and Domain Generalization Evaluation}
    \label{fig:generalizability}
  \end{minipage}
  \vspace{-0.25in}
\end{figure*}

\noindent \textit{Remark (One-time Data Collection and Pre-training Cost).} \textit{CARROT} requires 1.8 hours for data collection and about 7 minutes for agent training, which is acceptable compared to training-based methods such as ColBERT (10+ hours) and G-Retriever (6.2 hours). This is a one-time cost and can be reused across deployments, as the configuration agent generalizes to unseen domains and different LLMs (Figure~\ref{fig:generalizability}).

\subsubsection{\textbf{RQ3:} Scalability Evaluation} 
\begin{figure*}[!t]
  \centering
  \vspace{0in}
  \begin{minipage}[t]{0.35\textwidth}
    \centering
    \includegraphics[width=\textwidth]{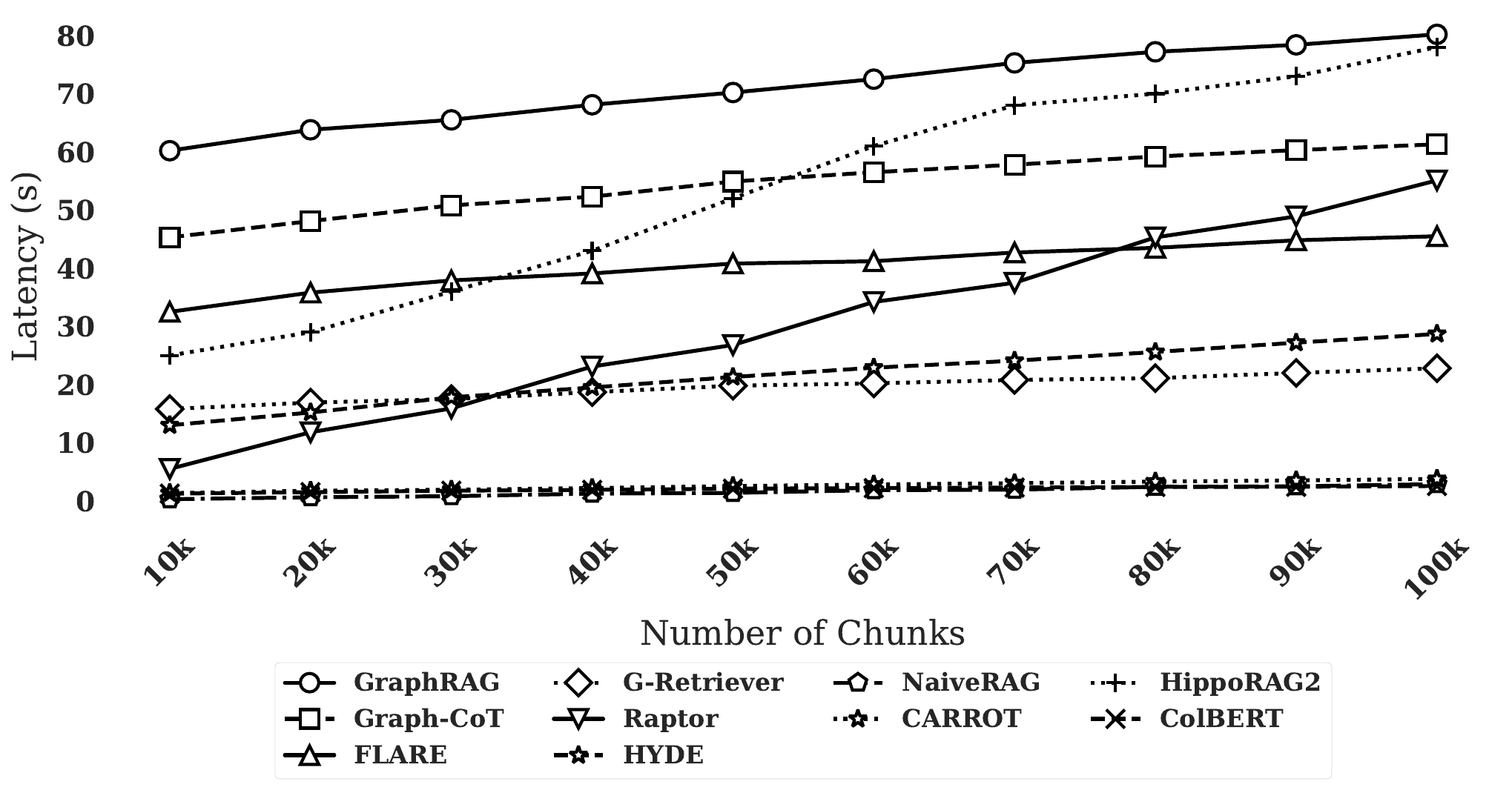}
    \captionof{figure}{\small{Scalability Comparison (WikiPassageQA)}}
    \label{fig:Scalability}
  \end{minipage}
  \hfill
  \begin{minipage}[t]{0.26\textwidth}
    \centering
    \includegraphics[width=0.85\textwidth]{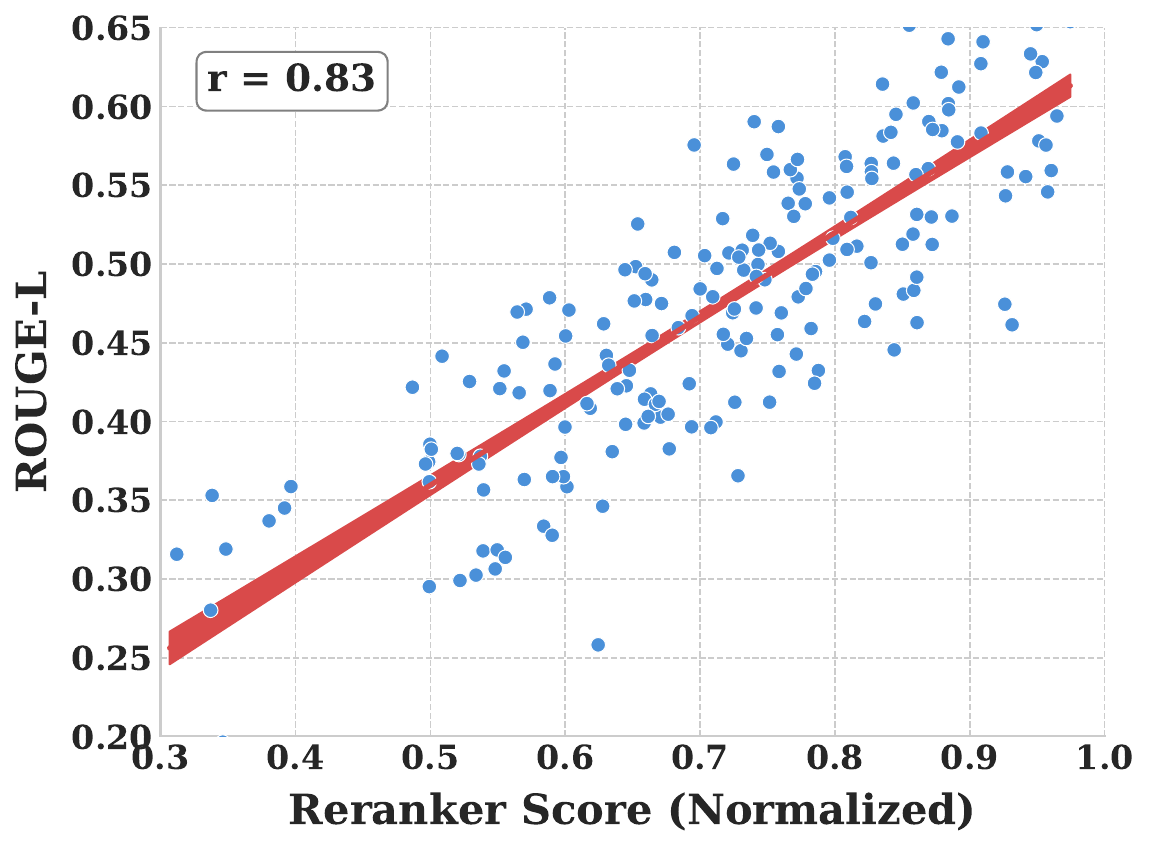}
    \captionof{figure}{\small{Reranker-Quality correlation}}
    \label{fig:reranker_correlation}
  \end{minipage}
  \hfill
  \begin{minipage}[t]{0.35\textwidth}
    \centering
    \includegraphics[width=\textwidth]{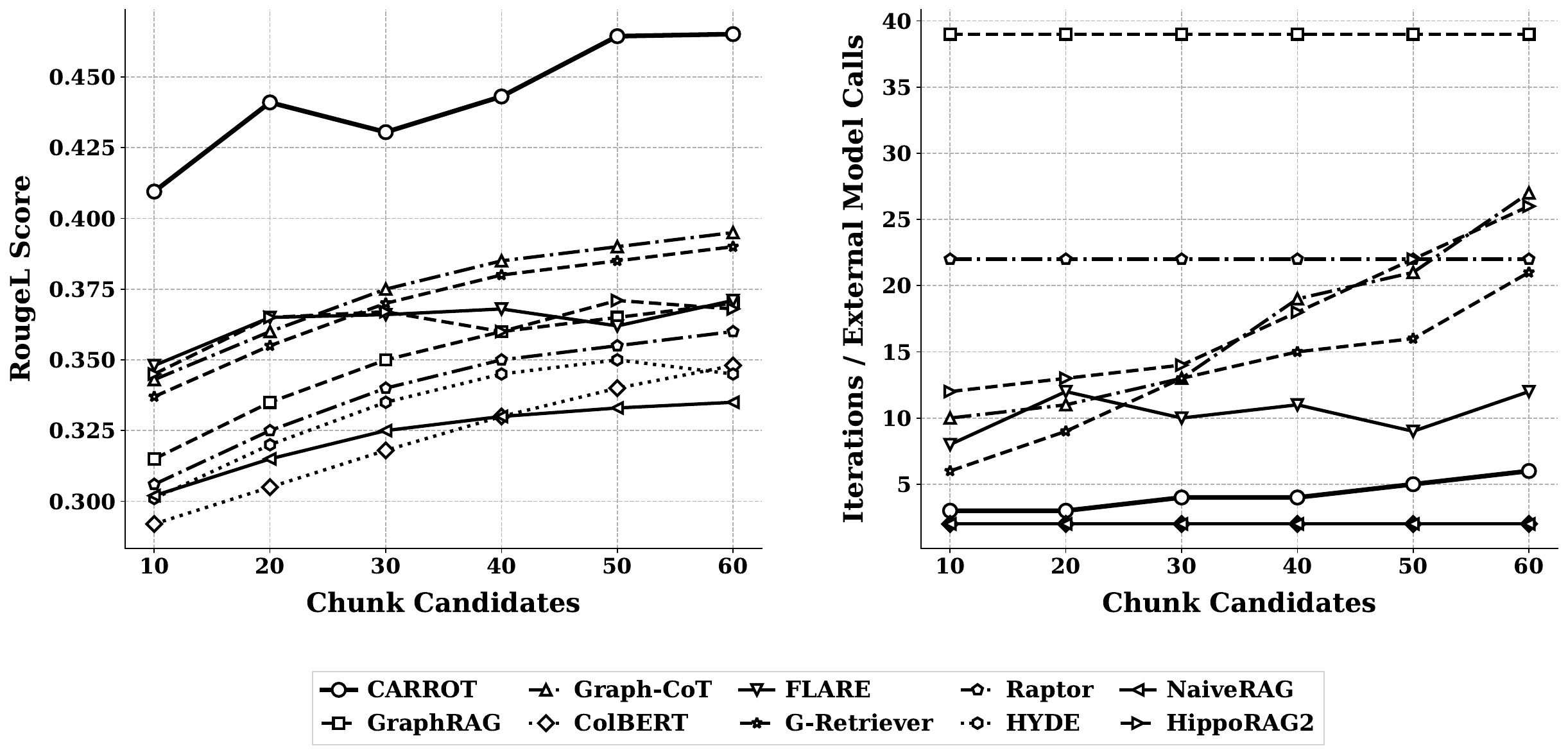}
    \vspace{-12pt}
    \captionof{figure}{\small{Quality/Efficiency vs. Candidates}}
    \label{fig:candidates_different}
  \end{minipage}
  \vspace{-0.1in}
\end{figure*}

To evaluate the scalability of \textit{CARROT}, we use the large-scale WikiPassageQA dataset and vary the number of chunks to assess performance under increasing data volumes. As shown in Figure~\ref{fig:Scalability}, we randomly sample between 10K and 100K chunks from the full dataset and measure the retrieval latency across different baselines. Despite the tenfold increase in data size, \textit{CARROT}'s latency rises by only about 10\%, demonstrating strong scalability in handling large-scale retrieval tasks. This stems from \textit{CARROT}'s ability to effectively prune the search space, maintaining a favorable balance between performance and computational overhead. In contrast, graph-based and online reasoning methods exhibit significantly higher overhead.

\subsubsection{\textbf{RQ4:} Effectiveness Evaluation on Different Budgets}
As demonstrated in Table~\ref{tab:budget}, we evaluate \textit{CARROT} under various budget constraints to assess its performance. Experiments conducted across the three datasets with budget limits of 1024, 2048, and 8192 tokens reveal that \textit{CARROT} consistently outperforms all baseline methods regardless of budget constraints. Notably, \textit{CARROT} consistently maintains average token consumption below designated budget thresholds. Furthermore, several baseline methods do not improve monotonically with larger budgets, empirically validating our observation in Challenge 2 that chunk utility does not follow a monotonic relationship with context length. In contrast, several baseline approaches either exceed the allocated budgets or consume them entirely, rendering such methods impractical for real-world deployment scenarios.

\noindent \textit{Remark.} For adaptable baselines like \textit{NaiveRAG} and \textit{HYDE}, we adapted these methods by limiting their token, stopping the addition of chunks when the output would exceed constraints. For graph-based methods such as \textit{GraphRAG} and \textit{HippoRAG2}, costs shown here reflect online token cost only; offline graph construction costs are discussed separately in Table~\ref{tab:training_comparsion}.

\subsection{RQ5: Ablation Study}
\subsubsection{Ablation Studies on Different Components.} Table \ref{tab:carrot_ablation} demonstrates that each component contributes substantially: removing the policy tree, configuration agent, or predicted parameters all cause significant performance decline. Notably, the parallel expansion mechanism enhances efficiency by reducing latency from 16.1s to 3.1s (80.7\% reduction). The latency breakdown (Table~\ref{tab:carrot_ablation} footnote) reveals that the Simulation phase requiring reranker inference dominates (48.4\%), while MCTS Tree Ops add only 9.7\% overhead, confirming that MCTS introduces acceptable computational cost for its quality gains. To further justify MCTS necessity, we compare five search strategies: (1) Top-k+JinaV2 (k=10) applies a fixed reranker to top-k candidates without exploring chunk combination orders, achieving R1=0.327; (2) Agent-Only extends the configuration agent to directly predict chunk combination order scores by adding a scoring head, bypassing MCTS search; it achieves the lowest quality because direct prediction cannot effectively consider chunk correlations; (3) Bandit~\cite{DBLP:journals/jmlr/UCB} employs arm-level exploration but neglects chunk combination order dependencies, achieving R1=0.362; (4) Beam Search~\cite{DBLP:conf/sigir/BeamQA} maintains top-k candidates per step yet remains limited by greedy local exploration, achieving R1=0.380; (5) \textit{CARROT} with MCTS systematically explores chunk order dependencies via tree-based global search, achieving R1=0.432, a 32.1\% gain over Top-k+JinaV2 with minimal overhead.

\subsubsection{Ablation Studies on Design Choices.}We conduct three ablation studies on key factors affecting RAG performance.

\textbf{(1) Validity of Reranker Scores as Simulation Reward.} In \textit{CARROT}'s policy tree search, the reranker assigns node benefits $W(\Phi_v)$ during the simulation phase to assess chunk combinations orders. Prior work~\cite{DBLP:journals/corr/jinarerankerv3,DBLP:conf/acl/lightrerank} has demonstrated high correlations between reranker scores and end-task quality; we empirically verify this property in our setting by sampling ${\sim}50\%$ queries from WikiPassageQA and enumerating diverse chunk combinations to collect reranker score, quality (ROUGE-L) pairs. As shown in Figure~\ref{fig:reranker_correlation}, we observe a Pearson correlation~\cite{obilor2018pearson} of $r{=}0.83$, confirming that reranker-assigned benefits reliably reflect downstream quality.

\textbf{(2) Impact of Chunk Candidate Counts.} As shown in Figure \ref{fig:candidates_different}, our analysis of chunk candidate counts on WikiPassageQA demonstrates that \textit{CARROT} achieves high response quality while requiring fewer iterations and thus fewer external model calls through its parallel expansion algorithm and policy tree search mechanism. Across different chunk candidate counts, several methods require more external model calls to select their optimal retrieval output, whereas \textit{CARROT} shows only a slight increase in external model call counts while maintaining them within an acceptable range and achieving response quality superior to baselines.

\textbf{(3) Sensitivity of Chunk Combination Order.} To isolate the effect of chunk combination order, we fix the chunk selection to top-5 chunks and vary only their ordering among four strategies: Random, Score-Descending (highest-first), Score-Ascending, and \textit{CARROT}. As shown in Figure~\ref{fig:order_sensitivity}, \textit{CARROT} outperforms the highest-first heuristic by 11--15\% relative gain. Figure~\ref{fig:order_sensitivity}(a) shows consistent ${\sim}11\%$ performance gap between best and worst orderings across Llama3-8B, DeepseekV3, and GPT-4o. Figure~\ref{fig:order_sensitivity}(b) reveals that multi-hop queries exhibit higher order sensitivity (12\% gap) than single-hop queries (11\%), suggesting complex reasoning benefits more from optimal ordering.

\subsubsection{Generalizability Evaluation}
To evaluate \textit{CARROT}'s generalizability, we conduct experiments across both domain and dataset dimensions. For domain-level generalization, we train the configuration agent on one domain's training set and evaluate on other domains' test sets using six distinct domains~\cite{DBLP:conf/acl/JinXZRZL0TWM024}, creating a 6×6 generalization matrix (where each cell represents the ROUGE-L score as shown in Figure~\ref{fig:generalizability}(a)). For dataset-level generalization, we train \textit{CARROT} on one dataset and test on entirely different datasets (to verify cross-dataset transferability, with ROUGE-L for WikiPassageQA and MARCO, F1 for HotpotQA presented in Figure~\ref{fig:generalizability}(b)). \textit{CARROT} exhibits robust generalization across both dimensions, maintaining competitive performance on training domains while achieving comparable results on unseen domains and datasets, demonstrating practical applicability.

\vspace{-0.05in}
\subsection{Case Study}
\vspace{-0.05in}
Figure~\ref{fig:case} presents qualitative examples comparing the retrieval performance of \textit{CARROT} against traditional baselines, illustrating how our method effectively addresses key challenges:
\begin{figure}
  \centering
  \includegraphics[width=0.98\linewidth]{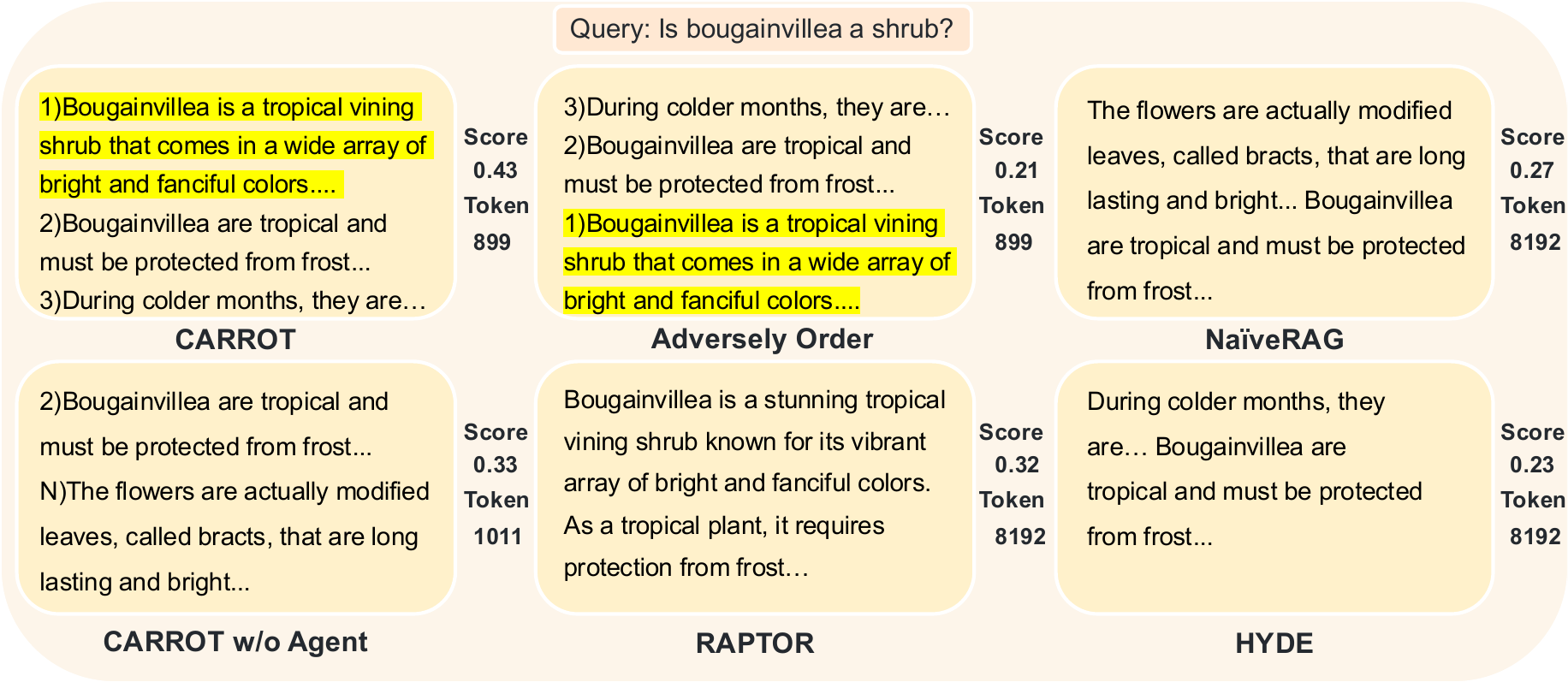}
  \vspace{-0.05in}
  \caption{Case Study for CARROT}
  \vspace{-0.3in}
  \label{fig:case}
\end{figure}
\noindent\textbf{(1) Importance of Chunk Combination Order.}
To evaluate the impact of chunk combination order, we test a deliberately reordered version of \textit{CARROT}'s output. The result shows a significant performance drop, with the ROUGE-1 score declining from 0.43 to 0.21, confirming that the chunk combination order is critical for RAG inference.

\noindent\textbf{(2) Non-monotonic Utility of Chunk Selection.}
We compare \textit{CARROT} with baselines such as \textit{RAPTOR}, \textit{NaiveRAG}, and \textit{HYDE}, all of which consume the full 8192 online input tokens but underperform relative to \textit{CARROT}, which uses only 899 online input tokens. For instance, \textit{NaiveRAG} relies on top-\textit{k} retrieval with reranking but greedily selects chunks without considering their correlations, often failing to capture semantically relevant context. When answering the query \textit{"Is bougainvillea a shrub?"}, it retrieves text containing matching terms but misses the essential taxonomic information. Meanwhile, \textit{HYDE} and \textit{RAPTOR} depend on costly LLM calls to generate hypotheses or summarize content, leading to inefficient token usage. In contrast, \textit{CARROT}'s adaptive chunk selection strategy retrieves concise, semantically aligned context, enabling more accurate and efficient LLM responses.

\noindent\textbf{(3) Handling Query Diversity.}
To assess robustness to diverse query types, we compare \textit{CARROT} with its variant \textit{CARROT w/o Agent}, which lacks the adaptive configuration agent. Without this component, the model struggles to identify optimal chunk combinations and often fails to retrieve the most relevant information, demonstrating the importance of the configuration agent in handling varied query intents.

\section{Conclusion}
Considering non-monotonic chunk utility, chunk correlation, and diverse query domains, we propose \textit{CARROT}, a learned cost-constrained retrieval optimization framework for RAG. We model chunk combination orders as a policy tree and apply MCTS to find the optimal combination within a cost constraint. A configuration agent predicts the best query configuration and reranker, while a parallel expansion strategy processes multiple nodes per iteration. Comprehensive experiments across three benchmark datasets demonstrate that \textit{CARROT} significantly outperforms state-of-the-art methods under constrained costs, achieving up to 30\% improvement while maintaining superior efficiency and scalability.

\section*{Acknowledgment}

This research is supported by Singapore MOE AcRF Tier-2 grant MOE-T2EP20223-0004. Wei Dong is supported by the National Research Foundation, Singapore, and the Cyber Security Agency of Singapore under the National Cybersecurity R\&D Programme and the CyberSG R\&D Programme Office (Award CRPO-GC3-NTU-001). Any opinions, findings, conclusions, or recommendations expressed in these materials are those of the author(s) and do not reflect the views of the National Research Foundation, Singapore, the Cyber Security Agency of Singapore, or the CyberSG R\&D Programme Office. We also thank Jingyi for helpful feedback.

\section*{AI-Generated Content Acknowledgment}

We use LLMs solely for language refinement, including grammar checking, sentence rephrasing, and improving clarity.

\bibliographystyle{IEEEtran}
    \bibliography{ref}
\end{document}